\documentclass[journal,10pt]{IEEEtran}
\usepackage{amsmath,amsfonts}
\usepackage{algorithmic}
\usepackage{algorithm}
\usepackage{bm}
\usepackage{array}  
\usepackage{comment}
\usepackage{threeparttable,multirow,booktabs}
\usepackage{xcolor,colortbl}
\definecolor{mygray}{gray}{0.88}
\usepackage{lineno}

\ifCLASSINFOpdf
  \usepackage[pdftex]{graphicx}
  \graphicspath{{figs/}}
  \DeclareGraphicsExtensions{.pdf}
\else
  \usepackage[dvips]{graphicx}
  \graphicspath{{eps/}}
  \DeclareGraphicsExtensions{.eps}
\fi

\usepackage{url,hyperref,bookmark}
\usepackage{pifont}
\usepackage{amssymb}
\usepackage{graphicx}
\usepackage{subcaption}

\usepackage{xspace}

\makeatletter
\DeclareRobustCommand\onedot{\futurelet\@let@token\@onedot}
\def\@onedot{\ifx\@let@token.\else.\null\fi\xspace}

\def\eg{\emph{e.g}\onedot} 
\def\ie{\emph{i.e}\onedot} 
 
\def\etc{\emph{etc}\onedot} 
 
\def\etal{\emph{et al}\onedot}
\def\I{\uppercase\expandafter{\romannumeral1}}
\def\II{\uppercase\expandafter{\romannumeral2}}
\makeatother

\hypersetup{draft}
\IEEEpubid{\begin{minipage}{\textwidth}\ \centering
		Copyright © 2025 IEEE. Personal use of this material is permitted. 
        However, permission to use this material for any other purposes must be obtained from the IEEE by sending an email to pubs-permissions@ieee.org.
\end{minipage}}

\begin{document}

\title{MarsQE: Semantic-Informed Quality Enhancement for Compressed Martian Image}

\author{
    Chengfeng~Liu,
    Mai~Xu,~\IEEEmembership{Senior~Member,~IEEE}, Qunliang~Xing,~\IEEEmembership{Graduate~Student~Member,~IEEE}
    and Xin~Zou
\thanks{
This work was supported by NSFC under Grants 62206011, 62250001 and 62231002, and Beijing Natural Science Foundation under Grant L223021. (Corresponding author: Mai Xu.)

Chengfeng Liu and Mai Xu are with the School of Electronic Information Engineering, Beihang University, Beijing 100191, China (e-mail: liuchengf@buaa.edu.cn; maixu@buaa.edu.cn).

Qunliang Xing is with the School of Electronic Information Engineering and Shen Yuan Honors College, Beihang University, Beijing 100191, China (e-mail: xingql@buaa.edu.cn).

Xin Zou is with the School of Electronic Information Engineering, Beihang University, Beijing 100191, China, and the Beijing Institute of Spacecraft System Engineering, Beijing 100191, China (e-mail: zouxin501@163.com).

}
}

\markboth{MarsQE: Semantic-Informed Quality Enhancement for Compressed Martian Image}%
{Liu \MakeLowercase{\textit{et al.}}: MarsQE: Semantic-Informed Quality Enhancement for Compressed Martian Image}

\maketitle

\begin{abstract}
Lossy image compression is essential for Mars exploration missions, due to the limited bandwidth between Earth and Mars. However, the compression may introduce visual artifacts that complicate the geological analysis of the Martian surface. Existing quality enhancement approaches, primarily designed for Earth images, fall short for Martian images due to a lack of consideration for the unique Martian semantics. In response to this challenge, we conduct an in-depth analysis of Martian images, yielding two key insights based on semantics: the presence of texture similarities and the compact nature of texture representations in Martian images. Inspired by these findings, we introduce MarsQE, an innovative, semantic-informed, two-phase quality enhancement approach specifically designed for Martian images. The first phase involves the semantic-based matching of texture-similar reference images, and the second phase enhances image quality by transferring texture patterns from these reference images to the compressed image. We also develop a post-enhancement network to further reduce compression artifacts and achieve superior compression quality. Our extensive experiments demonstrate that MarsQE significantly outperforms existing approaches for Earth images, establishing a new benchmark for the quality enhancement on Martian images.
The code is available at \url{https://github.com/keriphLiu/MarsQE}.

\end{abstract}


\section{Introduction}
\begin{figure*}[!t]
  \centering
  \includegraphics[width=.7\linewidth]{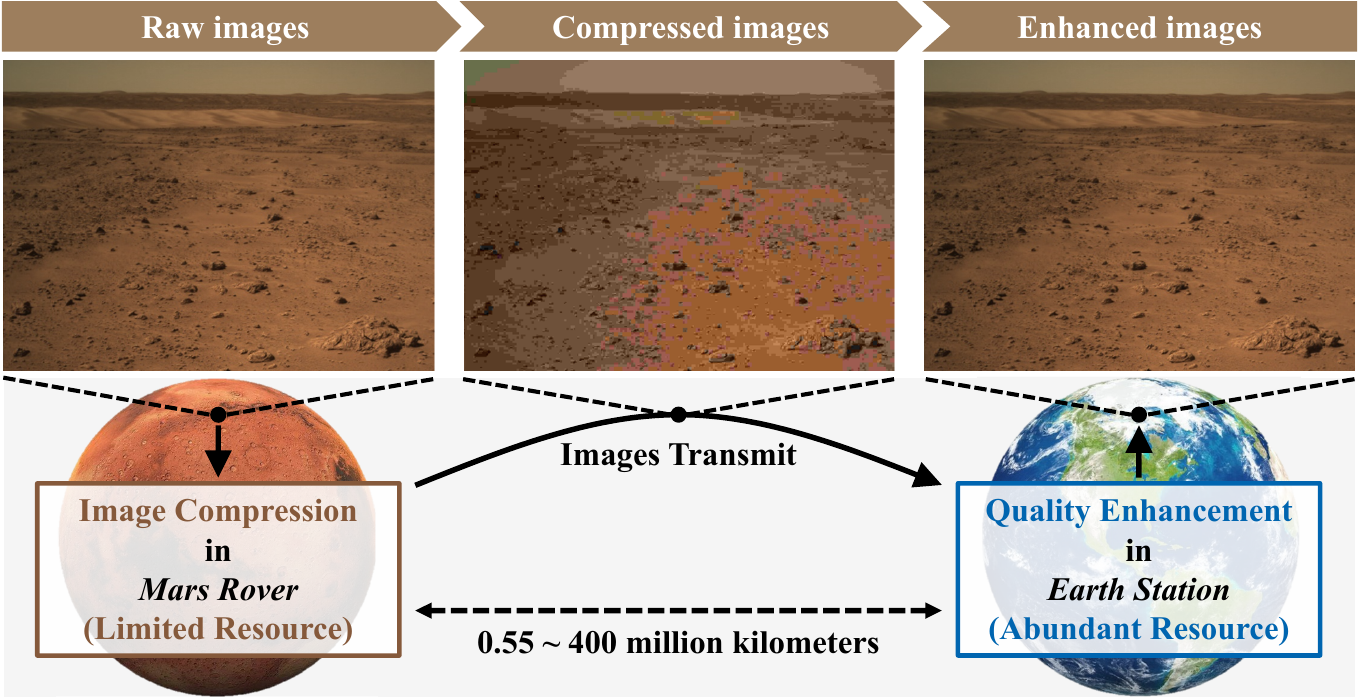}
  \caption{Motivation of our task.
  Raw Mars images are captured and compressed by Mars Rover.
  Then, compressed images are transmitted to Earth by Mars Orbiter.
  Finally, Ground Station received compressed images and enhance the quality of them to obtain enhanced images.}
  \label{fig1-fig}
  \vspace{-0.5em}
\end{figure*}

\IEEEPARstart{R}{ecently}, there has been growing interest in Mars exploration, due to the similarity between Mars and Earth~\cite{wan2020china,board2012vision}.
For example, from 2020 onwards, the NASA's Perseverance Mars rover and the Chinese TianWen-1 rover have successfully landed on Mars.
These rovers have captured a lot of valuable images, \eg, images of the Martian terrain, which are crucial for scientific research of Mars.
To address the challenge of images transmission from Mars to Earth given the limited bandwidth and energy, the lossy image compression techniques, such as Joint Photographic Experts Group (JPEG)~\cite{jpeg}, have been employed in the Mars Rover's encoder~\cite{malin2017mars,bell2017mars,meng2021high}.
However, the compression process inevitably incurs compression artifacts, \eg, ringing, blocking and blurring effects~\cite{1998shen}, which severely degrade the quality of Martian images.
Figure~\ref{fig1-fig} shows an example of compressed Martian image captured by the TianWen-1 rover.
As can be observed in this figure, the compression artifacts pose a significant challenge for geologic scientists in their analysis of the Martian surface.
To address this issue, a feasible solution is to leverage the abundant computational resources available in Earth stations to enhance the quality of compressed Martian images.

\begin{figure*}[!t]
  \centering
  \includegraphics[width=.9\linewidth]{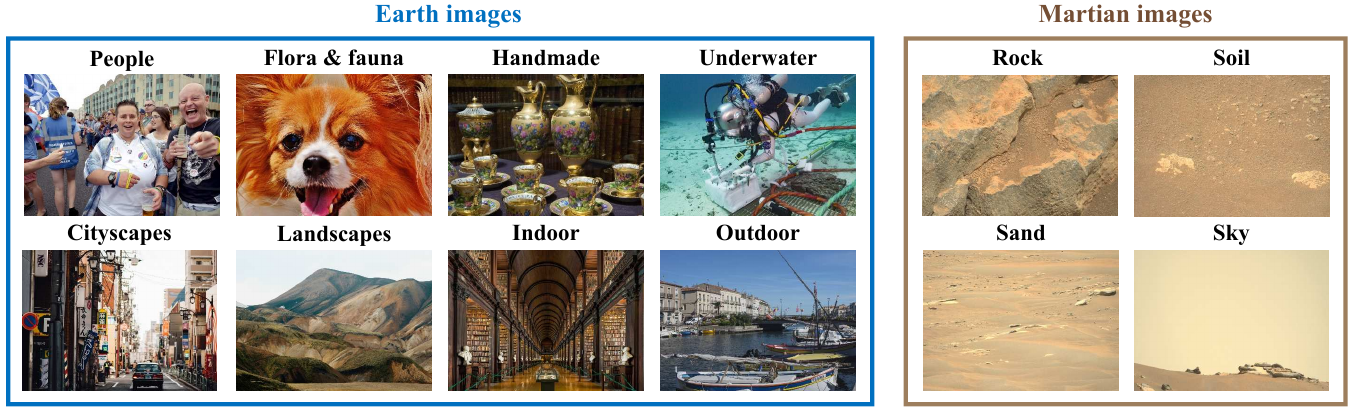}
  \caption{Examples of Earth images in DIV2K dataset and Martian images in the MIC dataset.}
  \label{fig2-fig}
  \vspace{-0.5em}
\end{figure*}

Meanwhile, the past few years have witnessed many approaches for quality enhancement on compressed images  captured on Earth~\cite{xiong1997deblocking,fan2000model,zhai2008efficient,ar-cnn,zhao2016reducing,dncnn,rbqe,daqe}.
Specifically, the Resource-efficient Blind Quality Enhancement (RBQE) approach~\cite{rbqe} is designed, trained and evaluated over the DIVerse 2K (DIV2K) dataset~\cite{div2k}, which includes eight diverse categories of content on Earth, \ie, people, flora \& fauna, handmade, cityscapes, landscapes, indoor, outdoor, and underwater.
However, Martian images are significantly different from Earth images, and possess much fewer semantics, as illustrated in Figure~\ref{fig2-fig}.
Two main characteristics of Martian images can be observed in this figure.
(1) Martian images display \textbf{notably greater inter-image and intra-image similarities} compared with Earth images (as noted in Finding~1).
(2) Martian images exhibit \textbf{more compact texture representation characterized by a limited number of semantic classes} compared with Earth images (as noted in Finding~2 and~3).
These characteristics are not applicable to Earth images, since the Earth images display a substantial diversity in texture patterns and semantics.
Consequently, the existing quality enhancement approaches designed for Earth images are inadequate for enhancing the quality of Martian images.

\IEEEpubidadjcol

In this paper, we propose a novel approach for enhancing the quality of compressed Martian images by considering two characteristics discussed above.
First, we analyze the Martian Image Compression (MIC) dataset~\cite{ding2022learning} to obtain the findings about these two characteristics.
Based on our findings, we propose a Martian images quality enhancement approach, named MarsQE.
Specifically, our MarsQE approach is designed with a two-stage quality enhancement network that utilizes notably greater pixel similarity within Martian images.
At the first stage, our network transfers texture features of pre-matched reference patches to restore texture details of compressed images in a patch-wise manner, such that the inter-image similarities can be utilized for quality enhancement.
Here, the Semantic-based Matching Module (SMM) is developed to search the reference patches across training images.
At the second stage, our network leverages similar texture features of distinct regions to reduce block effects and furthermore restore details, such that the intra-image similarities can be utilized.
Finally, we conduct extensive experiments on Martian image datasets to validate the state-of-the-art performance of our MarsQE approach.
In summary, the contributions of this paper are as follows:

\begin{itemize}
    \item [(1)] We conduct comprehensive analyses and present two key findings on Martian images, \ie, semantic-based similarity and compact texture representation.
    Theses findings are pivotal for quality enhancement and serve as the foundation of our MarsQE approach.
    \item [(2)] We propose a novel MarsQE approach for enhancing the quality of compressed Martian images. 
    In this task, the framework of MarsQE can effectively leverage the semantic-based texture similarity in Martian images through semantic-based reference matching and semantic-informed quality enhancement.
    \item [(3)] We conduct extensive experiments to validate the state-of-the-art performance of our MarsQE approach.
    We also verify its generalization capability on the Martian images from several Martian missions, \eg, Mars Science Laboratory (MSL) mission, without fine-tuning MarsQE, demonstrating the robustness of our approach.
\end{itemize}

\section{Related Works}\label{sec-related}

\subsection{Quality Enhancement for Earth Images}

Recently, extensive works ~\cite{dncnn,rbqe,ar-cnn,zhao2016reducing,2016guo,2016wang,kim2019pseudo,jin2020dual,mwcnn,cbdnet} have reached remarkable performance for enhancing the quality of Earth images, due to the booming development of Convolutional Neural Networks (CNNs).
Concretely, as the pioneer of CNN-based approaches, a shallow four-layer Artifacts Reduction Convolutional Neural Network (AR-CNN)~\cite{ar-cnn} was proposed to enhance the quality of JPEG-compressed images.
AR-CNN processively performs feature extraction, feature enhancement, feature mapping, and image reconstruction.
Notably, as a milestone of CNN-based approaches, Denoising Convolutional Neural Network (DnCNN)~\cite{dncnn} was proposed, which utilizes some effective techniques including the residual learning~\cite{2016he} and batch normalization~\cite{bn}.
With a 20-layer deep network, DnCNN can remove both Additive White Gaussian Noise (AWGN) and JPEG artifacts, surpassing most traditional approaches such as Block Matching and 3-D filtering (BM3D)~\cite{2007dabov}.
More recently, Xing \etal~\cite{rbqe} proposed a RBQE approach for the purpose of enhance quality in a resource-efficient manner.
By utilizing dynamic inference structure, RBQE approach can effectively and efficiently remove both compressed artifacts.
Guo \etal~\cite{cbdnet} proposed Convolutional Blind Denoising Network (CBDNet) to improve the generalization capability for real-world degradations.
The CBDNet was designed with an estimation subnetwork and a non-blind denoising subnetwork, such that effective blind quality enhancement can be achieved.
Later, the Defocus-Aware Quality Enhancement (DAQE) approach~\cite{daqe} was proposed to improve the quality of compressed images with a region-wise divide-and-conquer strategy, which utilizes image defocus to discern region-wise quality differences.

Furthermore, extensive research has been conducted on image enhancement under various adverse conditions, including dehazing, deraining, overexposure suppression, \etc.
Jiang \etal~\cite{jiang2024fmrnet} proposed FMRNet, which eliminates rain artifacts while preserving background textures in frequency space. 
Zhang \etal~\cite{zhang2024noise} developed Noise Self-Regression (NoiSER), a novel approach that utilizes pure Gaussian noise for training instead of task-specific data. 
Liu \etal~\cite{liu2022boths} proposed Boths, a lightweight approach for underwater image enhancement that integrates structural-detail feature interaction, 3-D attention learning, and dual-frequency loss functions. 
Liu \etal~\cite{liu2024towards} presented MM-UIE, which leverages a 6-D color operator to control the tone of enhanced images. 
Chi \etal~\cite{chi2024neural} proposed a dual-stage Fourier-based approach for faithful recovery of color and texture in shadow regions. 
Cui \etal~\cite{cui2025eenet} designed EENet, an efficient network for image dehazing through enhanced spatial-spectral learning.
Beyond synthetic scenes, several works have explored weakly-supervised or semi-supervised training for real-world image enhancement. 
Specifically, Su \etal~\cite{su2025real} proposed DNMPDT, a novel semi-supervised dehazing model for real scene image dehazing, which simultaneously utilizes real dataset and synthetic dataset to train a parameter-shared network.
Wang \etal~\cite{wang2025weakly} proposed a weakly supervised image dehazing model based on physical decomposition, which can improve the image dehazing effect while maintaining the authenticity of the enhanced image.

The aforementioned quality enhancement approaches for Earth images have employed datasets such as Berkeley Segmentation Data Set and Benchmarks 500 (BSDS500)~\cite{MartinFTM01} and DIV2K~\cite{div2k}.
These datasets comprise a large variety of contents with natural and artificial scenes, \eg, plants, man-made objects, cityscapes, \etc.
With these datasets, the aforementioned approaches can take advantage of the rich semantics in Earth images.
Although some recent approaches~\cite{xue2025fmtrack,xue2025target} incorporate auxiliary information from other domains, such as semantics, frequency, and other modalities, they are also designed specifically for Earth images.
In fact, different from Earth images, Martian images have limited semantics due to the simpler scene.
Therefore, we propose utilizing the limited-semantic characteristics to enhance the quality of Martian images.

\begin{figure}[!t]
  \centering
  \includegraphics[width=.9\linewidth]{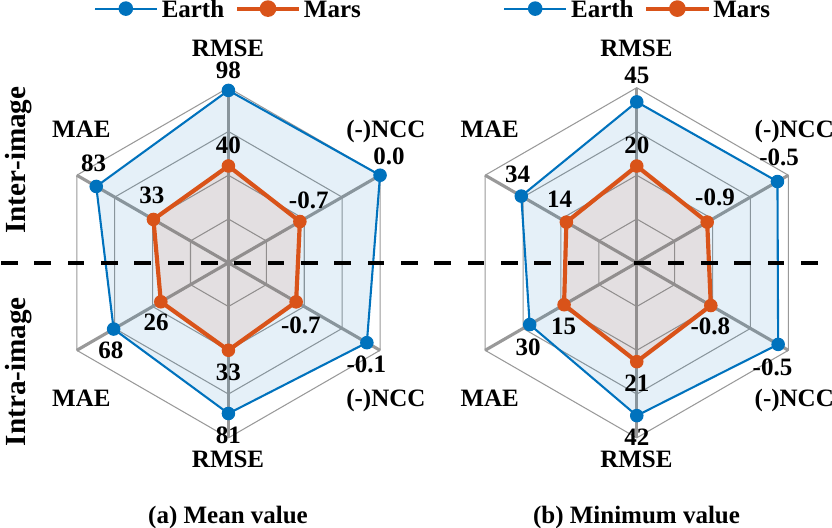}
  \caption{Evaluations of inter-image and intra-image similarities performed on Martian and Earth images.
  The measurement ranges for MAE and RMSE span from 0 to 100, while the range for NCC values extends from 0 to 1.
  To better present the results, NCC values are inverted and the range is halved for minimum value display.}
  \label{Similarity-Fig}
  \vspace{-1em}
\end{figure}

\begin{table}[!t]
    \caption{Evaluations of inter-image and intra-image similarities performed on other patch sizes. All the results are presented as mean/minimum values.}
    \label{tab-res-finding1-MainTab}
    \begin{subtable}[!t]{\linewidth}
    \centering
    \begin{tabular}{l|ccc}
    \toprule
    Size & \multicolumn{3}{c}{$128 \times 128$} \\
    \midrule
    Data & MIC & DIV2K & AID \\
    \midrule
    NCC & 0.73/0.87 & 0.01/0.64 & 0.05/0.58 \\
    MAE & 39.28/18.26 & 82.30/27.05 & 60.42/20.57 \\
    RMSE & 32.06/13.05 & 95.85/35.89 & 69.85/27.48 \\
    \toprule
    Size & \multicolumn{3}{c}{$64 \times 64$} \\
    \midrule
    Data & MIC & DIV2K & AID \\
    \midrule
    NCC & 0.75/0.90 & 0.01/0.71 & 0.06/0.66 \\
    MAE & 38.82/16.36 & 82.29/21.11 & 60.45/16.38 \\
    RMSE & 32.11/11.81 & 94.34/28.29 & 68.63/22.04 \\
    \bottomrule
    \end{tabular}
    \caption{Inter-image similarity analysis results.}
    \end{subtable} 
    \hfill
    \begin{subtable}[!t]{\linewidth}
    \centering
    \begin{tabular}{l|ccc}
    \toprule
    Size & \multicolumn{3}{c}{$128 \times 128$} \\
    \midrule
    Data & MIC & DIV2K & AID \\
    \midrule
    NCC & 0.75/0.85 & 0.12/0.55 & 0.17/0.42 \\
    MAE & 25.16/15.58 & 67.53/30.22 & 38.10/24.20 \\
    RMSE & 32.23/21.28 & 81.29/41.25 & 47.75/32.71 \\
    \toprule
    Size & \multicolumn{3}{c}{$64 \times 64$} \\
    \midrule
    Data & MIC & DIV2K & AID \\
    \midrule
    NCC & 0.77/0.87 & 0.14/0.66 & 0.19/0.56 \\
    MAE & 25.29/13.97 & 67.20/22.56 & 38.55/18.30 \\
    RMSE & 31.88/18.98 & 79.16/31.03 & 46.95/24.83 \\
    \bottomrule
    \end{tabular}
    \caption{Intra-image similarity analysis results.}
    \end{subtable}
    \vspace{-1.5em}
\end{table}

\subsection{Vision Works for Martian Images}

Some vision works have been conducted on Martian images to assist Mars exploration.
These vision works mainly focus on high-level vision tasks~\cite{ai4mars,deeplabv3_plus,panambur2022self,MarsTransformer}, including terrain classification and semantic segmentation.
Specifically, Michael \etal~\cite{ai4mars} employed a deep CNN architecture, called DeepLabV3+, for semantic segmentation on Martian terrain images.
The classification accuracy achieves over 96\%, showing the potential of auto-driving Mars rovers using deep-learning.
To make a step forward, Hu \etal~\cite{deeplabv3_plus} presented a hybrid network (DeepLabV3+/EfficientNet) to further improve the classification accuracy, such that the different areas of Mars can be judged.
From the perspective of geological taxonomy, Tejas \etal~\cite{panambur2022self} developed a self-supervised deep clustering algorithm to support rapid and robust terrain categorization.
By filtering data with high confidence of labels, Gagan \etal~\cite{MarsTransformer} designed SegFormer, a semantic segmentation framework utilizing vision transformer, for segmenting and identifying the terrain images of Mars.
In this way, the SegFormer can effectively segment terrain into three categories, \ie, sand, soil, and rock.
In addition to the high-level vision works, a few works focus on low-level vision tasks, such as image super-resolution and image compression~\cite{wang2021mars_sr,ding2022learning}.
Wang~\etal~\cite{wang2021mars_sr} addressed the challenge of reconstructing high-resolution Martian images from low-resolution ones by estimating the blur kernels.
Specifically, they proposed a novel degradation network specifically designed for Martian images.
Ding \etal~\cite{ding2022learning} proposed a deep Martian image compression network with a non-local block to explore both local and non-local dependencies among Martian image patches.

To the best of our knowledge, no prior work has been dedicated to the vital task of enhancing the quality of Martian images.
This becomes particularly significant for geologic analysis and science popularization, since Martian images invariably undergo compression for transmission, leading to the presence of noticeable artifacts~\cite{malin2017mars}.
Therefore, this paper makes the pioneering effort of enhancing the quality of Martian images.
Moreover, by leveraging the distinct and limited semantics inherent to Martian images, this paper effectively enhances the quality of Martian images, which may inspire further advancement in the domain of Martian vision processing.

\section{Findings}
This section presents our findings on spatial similarity and texture characteristics of Martian images in comparison with Earth images.
We first investigate the inter-image and intra-image similarities inherent in Martian images, which notably surpass those in Earth images.
Then, we investigate the texture patterns of Martian images, and find that their regions can be categorized into a few texture-distinct classes based on semantics.
These findings motivate our quality enhancement approach for Martian images, in particular the reference-based and semantic-informed quality enhancement as introduced in the next section.

\textbf{Finding 1:}
Martian images display significantly higher levels of both inter-image and intra-image similarity in comparison to Earth images.

\textbf{Analysis:}
Our analysis is based on the Martian images from the MIC dataset~\cite{ding2022learning} and the Earth images from the DIV2K dataset~\cite{div2k}.
To assess \textit{inter-image similarity}, we partition all images into non-overlapping patches with sizes of $256 \times 256$;
Subsequently, we pair patches from different images and assess their similarity using Mean Absolute Error (MAE), Root Mean Squared Error (RMSE), and Normalized Correlation Coefficient (NCC).
For the evaluation of \textit{intra-image similarity}, we pair patches from the same image and assess them using the same similarity metrics.

The evaluation results are shown in Figure~\ref{Similarity-Fig}.
Note that smaller MAE/RMSE values and higher NCC values signify a higher degree of similarity between two image patches.
As seen in Figure~\ref{Similarity-Fig}, Martian image patches exhibit significantly higher similarity compared to Earth images, for both inter-image and intra-image comparisons.
For instance, the average inter-image RMSE for Martian images stands at 40, representing only 40.8\% of the corresponding value for Earth images (\ie, 98).
Similarly, the average intra-image RMSE for Martian images is 33, accounting for just 40.7\% of the corresponding value for Earth images (\ie, 81).
We also conduct analysis on patches with sizes of $128 \times 128$ and $64 \times 64$.
As shown in Table~\ref{tab-res-finding1-MainTab}, similar trends are observed across other metrics and patch sizes.
Beyond Earth images from DIV2K, we extend our experiments to remote sensing images from the Aerial Image Dataset (AID)~\cite{AID}.
Table~\ref{tab-res-finding1-MainTab} further confirms consistent performance trends when replacing natural images with remote sensing images.
In conclusion, it is evident that Martian images exhibit markedly greater inter-image and intra-image similarity, compared with Earth images. 

Moreover, we employ mathematical formulations to demonstrate these two characteristics of similarity.
Given the patch sets extracted from their respective image datasets, \ie, the Martian patch set $\{\{\mathbf{P}^{M}_{i,j}\}_{j=1}^{N_p}\}_{i=1}^{N_m}$ and the Earth patch set $\{\{\mathbf{P}^{E}_{i,j}\}_{j=1}^{N_p}\}_{i=1}^{N_e}$, where $\mathbf{P}^{M}$ denotes the Martian image patch, $\mathbf{P}^{E}$ denotes the Earth image patch, ${N_p}$ denotes the number of patches per image, ${N_m}$ represents the number of images in the Martian dataset, and ${N_e}$ denotes the number of images in the Earth dataset.
The phenomenon that Martian images display notably greater inter-image similarities compared with Earth images can be defined as:
\begin{equation}
    \mathbb{E}_{i \neq p}[\text{sim}(\mathbf{P}^{M}_{i,j},\mathbf{P}^{M}_{p,q})] > \mathbb{E}_{i \neq p}[\text{sim}(\mathbf{P}^{E}_{i,j},\mathbf{P}^{E}_{p,q})],
\end{equation}
where $\text{sim}$ denotes the similarity measure.
The phenomenon that Martian images display notably greater intra-image similarities compared with Earth images can be defined as:
\begin{equation}
    \mathbb{E}_{j \neq k}[\text{sim}(\mathbf{P}^{M}_{i,j},\mathbf{P}^{M}_{i,k})] > \mathbb{E}_{j \neq k}[\text{sim}(\mathbf{P}^{E}_{i,j},\mathbf{P}^{E}_{i,k})].
\end{equation}
This concludes the analysis of Finding 1.

\begin{figure}
  \centering
  \begin{subfigure}[c]{.85\linewidth}
    \includegraphics[width=\linewidth]{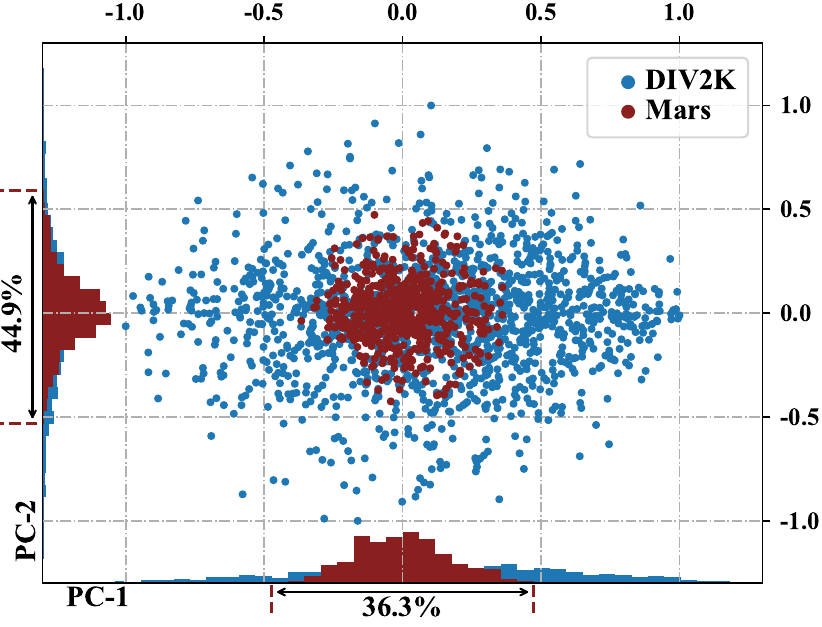}
    \caption{Texture representations of Martian and DIV2K images.}
    \label{Finding2_DIV2K}
  \end{subfigure}
  \hfill
  \begin{subfigure}[c]{.85\linewidth}
    \includegraphics[width=\linewidth]{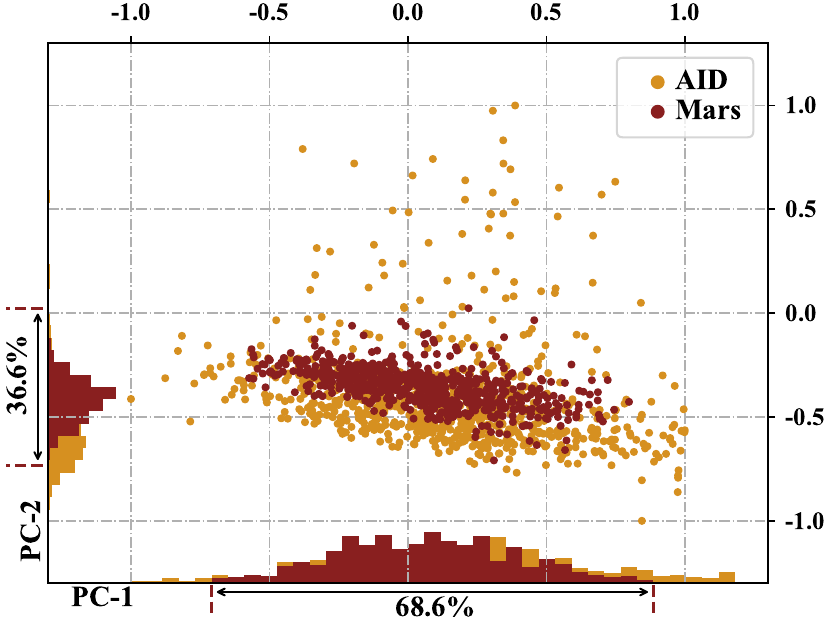}
    \caption{Texture representations of Martian and AID images.}
    \label{Finding2_AID}
  \end{subfigure}
  \caption{Texture representations of Martian and Earth images.
  The two dimensions denoted by PC-1 and PC-2 measure the intensity of the two most significant and representative textural features extracted by PCA.
  Their values are normalized to the range $[-1, 1]$ according to the values of Earth images (from DIV2K dataset or AID).
  In addition, the frequency histogram of the texture representation is also provided.}
  \label{fig_res_finding2}
\end{figure}

\textbf{Finding 2:}
Compared with Earth images, Martian images exhibit a higher degree of compactness in their texture representation.

\textbf{Analysis:}
For both Martian and Earth images, we employ the widely-recognized texture descriptor, \ie, Local Binary Pattern (LBP)~\cite{LBP}, to capture texture patterns from image patches.
Subsequently, we project the high-dimensional LBP textures to a two-dimensional feature space using Principal Component Analysis (PCA)~\cite{PCA}, which extracts the two most significant and representative textural features, denoted by PC-1 and PC-2.
Consequently, this two-dimensional feature space serves as a powerful representation for Martian and Earth texture patterns.
As depicted in Figure~\ref{fig_res_finding2}, the texture representation of Martian images is noticeably more concentrated compared to that of both natural images in the DIV2K dataset and remote sensing images in the AID.
Furthermore, we provide a frequency histogram of the texture representation in this figure.
Specifically, the Martian texture representation exhibits only 36.3\% and 44.9\% of the range observed in DIV2K's texture representation along the PC-1 and PC-2 dimensions, respectively.
These results indicate that the texture representations of Martian images are more compact than those of Earth images.

Moreover, we employ mathematical formulations to demonstrate the compactness of the texture representation. 
Given the covariance matrices $\Sigma^M$ and $\Sigma^E$ for the Martian and Earth texture representations, a more compact texture representation is indicated by a smaller trace of the covariance matrix, formulated as:
\begin{equation}
    \text{Tr}(\Sigma^M) < \text{Tr}(\Sigma^E).
\end{equation}
This concludes the analysis of Finding 2.

\textbf{Finding 3:}
By leveraging the semantic classification, the regions of Martian images can be categorized into a limited number of texture-distinct semantic classes.

\begin{figure}[!t]
  \centering
  \includegraphics[width=1.\linewidth]{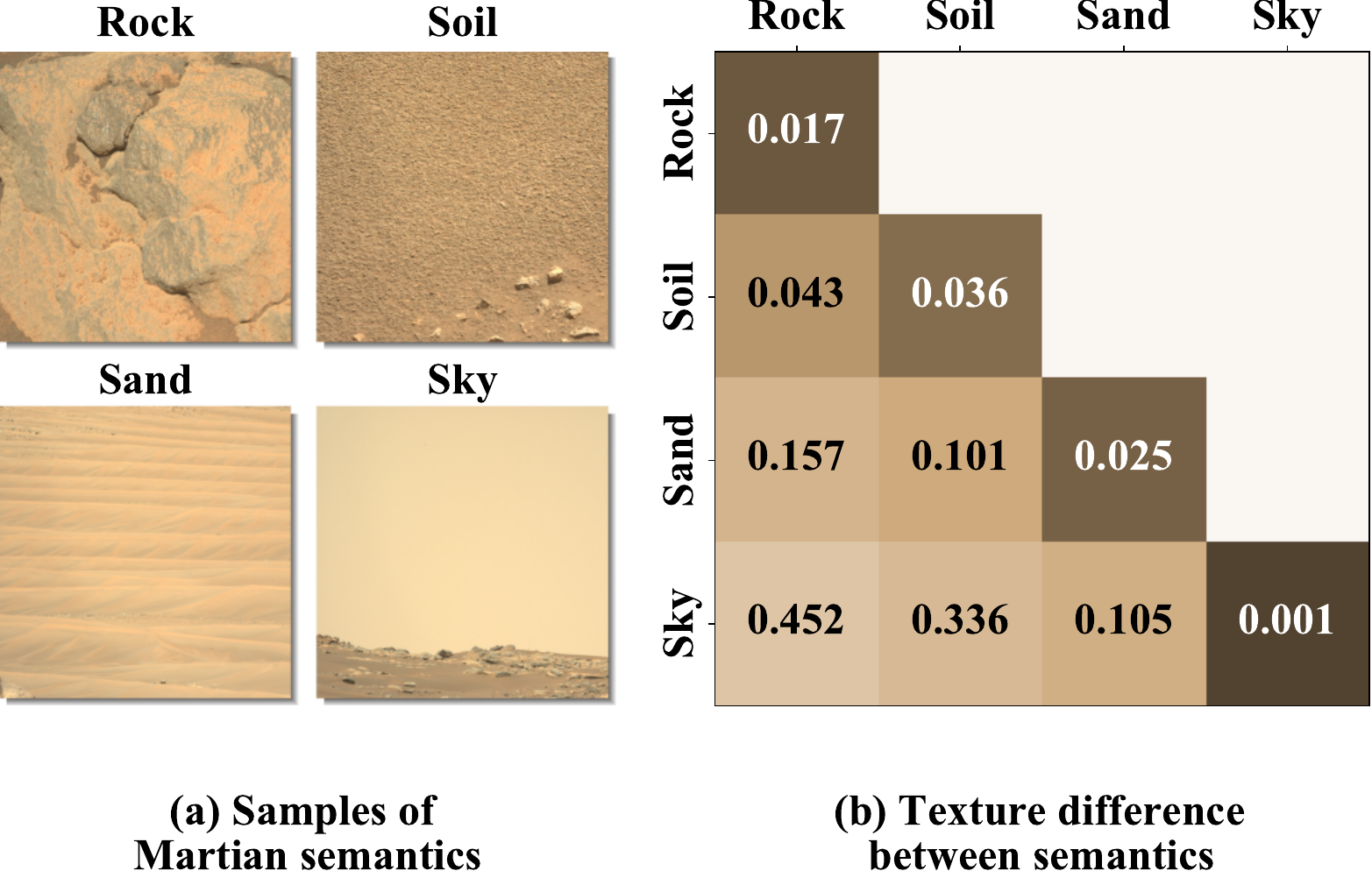}
  \caption{Martian images can be categorized into a few semantic classes that differentiate in texture patterns.
  The texture difference is measured by the JS divergence values between LBP features.}
  \label{Finding3-Fig}
  \vspace{-0.5em}
\end{figure}

\textbf{Analysis:}
Finding 2 highlights the simpler texture representation of Martian images compared to Earth images.
Building upon this insight of Finding 2, we delve deeper to explore the potential categorization of Martian regions into texture-distinct semantic classes.
Following existing works on Martian terrain segmentation~\cite{ai4mars,MarsTransformer}, we first categorize Martian regions into four semantic classes, \ie, sand, soil, rock, and sky, as illustrated in Figure~\ref{Finding3-Fig} (a).
Subsequently, we measure the average texture dissimilarity between patches belonging to the same or different semantic classes, by evaluating their Jensen–Shannon (JS) divergence of LBP patterns.
Note that a smaller JS divergence value indicates a higher similarity of the texture patterns between two patches.
Figure~\ref{Finding3-Fig} (b) illustrates that the JS divergence values between patches from different semantic classes are notably greater than those between patches from the same class. For instance, the average JS divergence value between two patches categorized as rock is 0.017, which is 96.2\% lower than the corresponding value between two patches from the sky and rock categories (\ie, 0.452).
Therefore, we can conclude that Martian images can be effectively represented by a small number of semantic classes that exhibit distinct texture patterns.
Finally, the analysis of Finding 3 is accomplished.

\begin{figure*}[!t]
    \centering
    \includegraphics[width=1.\linewidth]{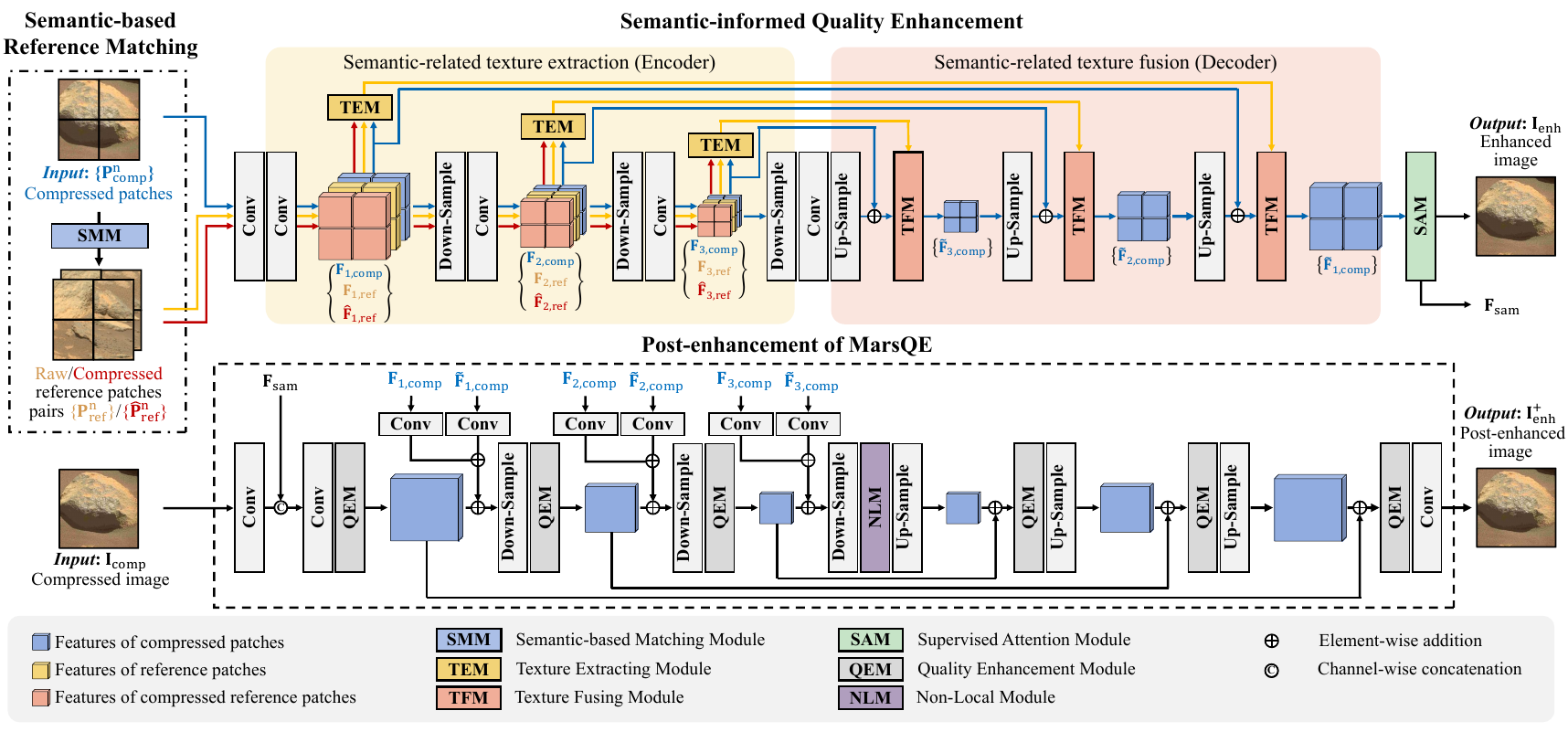}
    \caption{Framework of the proposed MarsQE approach. The level and division number of each network are set to three and four in the figure for illustration.}
    \label{fig-framework}
    \vspace{-0.5em}
\end{figure*}

\section{Proposed Approach}\label{sec-app}

In this section, we propose the MarsQE approach for enhancing the quality of compressed Martian images.
Leveraging the similarity among Martian images as observed in our findings, MarsQE consists of two primary steps: 1) Semantic-based reference matching (Section~\ref{sec-sec-SMM}) and 2) Semantic-informed quality enhancement (Section~\ref{sec-sec-MainNet}). 
Additionally, a post-enhancement network is proposed to further mitigate blocking artifacts and refine texture details (Section~\ref{sec-sec-ExtrcNet}). Figure~\ref{fig-framework} shows the overall framework of our MarsQE approach. More details are presented in the following.

\subsection{Semantic-based Reference Matching}\label{sec-sec-SMM}

Given the significant inter-image similarity in Martian images (as noted in Finding~1), MarsQE takes advantages of texture-similar reference images for quality enhancement.
As shown in the dot-dash box of Figure~\ref{fig-framework}, the Semantic-based Matching Module (SMM) is designed to efficiently obtain these reference images.

\textbf{Reference dictionary construction:}
Findings 2 and 3 reveal that Martian images can be efficiently categorized into a few texture-distinct semantic classes.
Inspired by this characteristic, SMM constructs a compact but comprehensive semantic-based dictionary.
Specifically, SMM first divides the raw images into non-overlapping $S \times S$ patches.
Then, these patches are categorized into $K$ semantic classes, utilizing Martian terrain segmentation~\cite{MarsTransformer}.
Finally, for the $k$-th semantic class where $k$ ranges from $1$ to $K$, SMM selects $N_k$ patches to construct a reference dictionary $\{\mathbf{R}^{k}_{i}\}_{i=1}^{N_k}$.
The corresponding compressed patches $\{\hat{\mathbf{R}}^{k}_{i}\}_{i=1}^{N_k}$ are also included for reference.

\textbf{Reference patch matching:}
SMM conducts patch-wise reference matching using the obtained reference dictionary.
Specifically, the input compressed image $\mathbf{I}_{\text{comp}}$ is divided into ${n}$ non-overlapping patches $\{\mathbf{P}^{n}_{\text{comp}}\}$.
In the Figure~\ref{fig-framework}, the input compressed image is divided into four patches for illustration, \ie, ${n=4}$.
Each compressed patch $\mathbf{P}_{\text{comp}}$ from $\{\mathbf{P}^{n}_{\text{comp}}\}$ is semantically classified and matched with the appropriate patches in the same-semantic-class dictionary.
The matching process is formulated as:
\begin{equation}
    n = \underset{i}{\operatorname{arg\,min}} \ 
    D_{\text{KL}} \Bigl(
    \operatorname{\text{LBP}} \bigl(
    \hat{\mathbf{R}}^{k}_{i} \bigr) \Vert \operatorname{\text{LBP}} \bigl( \mathbf{P}_{\text{comp}} \bigr) \Bigr).
\end{equation}
In the above formulation, $D_{\text{KL}} (x \Vert y)$ denotes the Kullback-Leibler (KL) divergence~\cite{kld} between vectors $x$ and $y$;
$\operatorname{\text{LBP}} (\cdot)$ denotes the LBP~\cite{LBP} texture descriptor.
As a result, the semantically consistent reference patches $\mathbf{R}^{k}_{n}$ and $\hat{\mathbf{R}}^{k}_{n}$ are selected for each compressed patch $\mathbf{P}_{\text{comp}}$ as $\mathbf{P}_{\text{ref}}$ and $\hat{\mathbf{P}}_{\text{ref}}$, respectively.

\begin{figure*}
  \centering
  \begin{subfigure}[c]{0.3990\linewidth}
    \includegraphics[width=\linewidth]{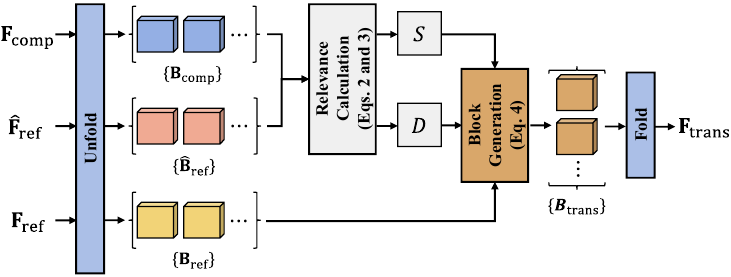}
    \caption{Texture Extracting Module (TEM).}
    \label{TEM}
  \end{subfigure}
  \hfill
  \begin{subfigure}[c]{0.2722\linewidth}
    \includegraphics[width=\linewidth]{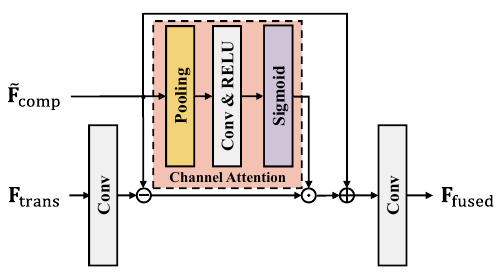}
    \caption{Texture Fusing Module (TFM).}
    \label{TFM}
  \end{subfigure}
  \hfill
    \begin{subfigure}[c]{0.3088\linewidth}
    \includegraphics[width=\linewidth]{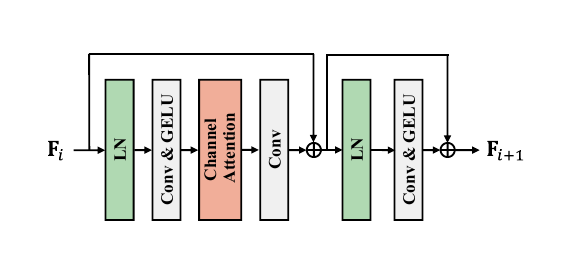}
    \caption{Quality Enhancement Module (QEM).}
    \label{QEM}
  \end{subfigure}
  \caption{Module architectures adopted in MarsQE.}
  \label{fig_modules}
  \vspace{-1em}
\end{figure*}

\subsection{Semantic-informed Quality Enhancement}\label{sec-sec-MainNet}

Given the obtained raw and compressed reference patches, the MarsQE approach performs semantic-informed quality enhancement in a patch-wise manner.
According to Findings 2 and 3, patches from the same semantic classes exhibit similar texture patterns.
Thus, MarsQE uses semantic-related textures from reference patches to guide the enhancement of input patches.
Specifically, the overall network is designed with a $L$-level encoder-decoder architecture.
The network initially extracts transferable features with semantic-related textures at various scales, and then fuses these features with decoded features to restore texture details.
Note that the input patches $\{\mathbf{P}^{n}_{\text{comp}}\}$ are processed in a patch-wise manner, yielding encoder features $\{\mathbf{F}^{n}_{l,\text{comp}}\}$ and decoder features $\{\widetilde{\mathbf{F}}^{n}_{l,\text{comp}}\}$.
$l$ denotes to the level of the feature.
$\{\mathbf{F}^{n}_{l,\text{comp}}\}$ and $\{\widetilde{\mathbf{F}}^{n}_{l,\text{comp}}\}$ are spatially recombined to the input image size, resulting in $\mathbf{F}_{l,\text{comp}}$ and $\widetilde{\mathbf{F}}_{l,\text{comp}}$.
The same process is applied to raw and compressed reference patches, \ie, $\{\mathbf{P}^{n}_{\text{ref}}\}$ and $\{\hat{\mathbf{P}}^{n}_{\text{ref}}\}$.
Finally, the fused features are enhanced through the Supervised Attention Module (SAM)~\cite{mprnet}, allowing for efficient feature propagation to the post-enhancement network.

\textbf{Semantic-related texture extraction:}
In fact, reference patches are rich in semantic-related texture details, crucial for enhancing the quality of input compressed patches.
As depicted in the yellow box of Figure~\ref{fig-framework}, the network employs an encoder to extract multi-scale semantic-related textures from the compressed input, raw reference, and compressed reference patches.
The extraction flows for these patches are represented by blue, yellow, and red lines, respectively.
Texture Extracting Modules (TEMs) are integral to this process, designed to extract transferable features at each scale.
Consider one input patch for example as shown in Figure~\ref{TEM}.
The encoded features of $\mathbf{P}_{\text{comp}}$, $\hat{\mathbf{P}}_{\text{ref}}$, and $\mathbf{P}_{\text{ref}}$, denoted as $\mathbf{F}_{\text{comp}}$, $\hat{\mathbf{F}}_{\text{ref}}$, and $\mathbf{F}_{\text{ref}}$, are fed into TEM as query, key and value.
TEM unfolds these features into non-overlapping $d \times d$ feature blocks, resulting in $\{\mathbf{B}_{\text{comp}}\}$, $\{\hat{\mathbf{B}}_{\text{ref}}\}$, and $\{\mathbf{B}_{\text{ref}}\}$.
For each block $\mathbf{B}_{\text{comp}}^{i}$, TEM identifies the most relevant block in $\{\hat{\mathbf{B}}_{\text{ref}}\}$ as follows,
\begin{gather}
    {D}_{i} = \underset{j}{\operatorname{arg\,max} } \left\langle \frac{\mathbf{B}_{\text{comp}}^{i}}{\Vert \mathbf{B}_{\text{comp}}^{i} \Vert} , \frac{\hat{\mathbf{B}}_{\text{ref}}^{j}}{\Vert \hat{\mathbf{B}}_{\text{ref}}^{j} \Vert} \right\rangle,\\
    {S}_{i} = \underset{j}{\max } \left\langle \frac{\mathbf{B}_{\text{comp}}^{i}}{\Vert \mathbf{B}_{\text{comp}}^{i} \Vert} , \frac{\hat{\mathbf{B}}_{\text{ref}}^{j}}{\Vert \hat{\mathbf{B}}_{\text{ref}}^{j} \Vert} \right\rangle,
\end{gather}
where ${D}_{i}$ indicates the index and ${S}_{i}$ represents the similarity score.
The transferable feature blocks $\{\mathbf{B}_{\text{trans}}\}$ are then enhanced from $\{\mathbf{B}_{\text{ref}}\}$:
\begin{equation}
    {\mathbf{B}}_{\text{trans}}^{i} = \mathbf{B}_{\text{ref}}^{D_{i}} \times S_{i}.
\end{equation}
Finally, we aggregate $\{\mathbf{B}_{\text{trans}}^{n}\}$ to form the transferable features $\mathbf{F}_{\text{trans}}$, thereby extracting the semantic-related textures.

\textbf{Semantic-related texture fusion:}
After extracting the semantic-related textures, the learnable decoder employs these textures to enrich the decoded features.
This process is visually detailed in the red box of Figure~\ref{fig-framework}.
At each decoder level, the decoded features are up-sampled and combined with corresponding encoded features via skip connections.
These features are then fused with the extracted transferable features using the Texture Fusing Module (TFM), as depicted in Figure~\ref{TFM}.
TFM initially calculates the residual $\mathbf{F}_{\text{res}}$, which represents the difference between transferable features $\mathbf{F}_{\text{trans}}$ and decoded features $\widetilde{\mathbf{F}}_{\text{comp}}$.
Subsequently, TFM generates attention maps to filter informative features from $\mathbf{F}_{\text{res}}$ for subsequent fusion.
The process culminates with the fusion of $\widetilde{\mathbf{F}}_{\text{comp}}$ and the filtered features, output through a convolution layer, resulting in the fused features $\mathbf{F}_{\text{fused}}$.
Mathematically, the above process is represented by
\begin{equation}
    \left\{
    \begin{aligned}
    &\mathbf{F}_{\text{res}} = C_{3 \times 3} \bigl( \mathbf{F}_{\text{trans}} \bigr) - \widetilde{\mathbf{F}}_{\text{comp}}, \\
    &\mathbf{F}_{\text{fused}} = C_{3 \times 3} \bigl(
    \widetilde{\mathbf{F}}_{\text{comp}} + \bigl(
     \mathbf{F}_{\text{res}} \odot \operatorname{\text{CA}}
    \bigl( \widetilde{\mathbf{F}}_{\text{comp}} 
            \bigr) 
        \bigr) 
    \bigr).
    \end{aligned}
    \right.
\end{equation}
In the above, $C_{3 \times 3} ( \cdot )$ represents a convolution layer with a kernel size of $3 \times 3$, and $\operatorname{\text{CA}} ( \cdot )$ signifies a channel attention layer~\cite{ChannelAttention}.
At the output side of the network, the final decoded features $\widetilde{\mathbf{F}}_{1,\text{comp}}$ are spatially recombined to the input image size and fed into SAM~\cite{mprnet} to generate an enhanced image $\mathbf{I}_{\text{enh}}$.
This way, the semantic-related textures from reference patches are effectively utilized to produce a semantically enriched enhanced image.

\subsection{Post-enhancement of MarsQE}\label{sec-sec-ExtrcNet}

The main network of MarsQE has produced a high-quality enhanced image.
In addition, we develop a post-enhancement network to mitigate blocking effects caused by patch-wise enhancement, such that semantic details can be enriched.
It is also designed with an encoder-decoder architecture. In the encoder stage, the features from the main network are combined with the encoded features of the input compressed image $\mathbf{I}_\text{comp}$.
These features are further refined at each level by the Quality Enhancement Module (QEM), as illustrated in Figure~\ref{QEM}.
QEM is mainly composed of Layer Normalization (LN)~\cite{LayerNorm}, Gaussian Error Linear Unit (GELU)~\cite{GELU}, and Channel Attention (CA)~\cite{ChannelAttention}.

To exploit the intra-image similarity in Martian images (highlighted in Finding~1), the network incorporates the Non-Local Module (NLM) for each compressed image.
Given the texture-distinct semantic classes obtained in Section~\ref{sec-sec-SMM}, we apply NLMs within each semantic class, significantly accelerating the post-enhancement process.
The functionality of NLM is mathematically represented as
\begin{equation}
    \left\{
    \begin{aligned}
    &\mathbf{X}_{i} =
    C_{3 \times 3} \bigl(  
    L_{\text{NLA}} \bigl( 
    L_{\text{LN}} \bigl(
    \mathbf{F}_{i} \bigr) \bigr) \bigr) + \mathbf{F}_{i}, \\
    &\mathbf{F}_{i+1} = C_{1 \times 1} \bigl(
    L_{\text{GELU}} \bigl(
    C_{1 \times 1} \bigl(
    L_{\text{LN}} \bigl(
     \mathbf{X}_{i} 
                \bigr) 
            \bigr) 
        \bigr) 
    \bigr) + \mathbf{X}_{i}.
    \end{aligned}
    \right.
\end{equation}
Here, $\mathbf{F}_{i}$ and $\mathbf{F}_{i+1}$ represent the input and output features of NLM, respectively.
The notations $C_{1 \times 1} ( \cdot )$, $L_{\text{NLA}} ( \cdot )$, $L_{\text{LN}} ( \cdot )$, and $L_{\text{GELU}} ( \cdot )$ correspond to a convolution layer with a kernel size of $1 \times 1$, a non-local attention layer, a LN layer, and a GELU activation layer, respectively.
The features produced by NLM are then fed into the decoder, resulting in a final enhanced image $\mathbf{I}_{\text{enh}}^{\text{+}}$, exhibiting higher quality than $\mathbf{I}_{\text{enh}}$.
The overall framework of MarsQE is trained in an end-to-end manner with the loss function of Mean Squared Error (MSE) between the ground truth image $\mathbf{I}_{\text{raw}}$ and two enhanced images:
\begin{equation}    
    \mathcal{L}_{\text{MarsQE}} = \Vert \mathbf{I}_{\text{enh}} - \mathbf{I}_{\text{raw}} \Vert_2^2 + \lambda \Vert \mathbf{I}_{\text{enh}}^{\text{+}} - \mathbf{I}_{\text{raw}} \Vert_2^2.
\end{equation}
In summary, our MarsQE approach effectively enhances the quality of compressed Martian images by leveraging semantic-related intra- and inter-image similarities.

\section{Experiment}

\subsection{Experimental Setup}\label{sec-sec-setup}

In this section, we present details about the datasets, hyper-parameters, and training strategy of our experiments for evaluating the effectiveness of MarsQE.

\begin{table}
    \caption{Datasets adopted in our experiments.
    The usage of these datasets, their size, the existence of ground-truth, and which rover they came from are indicated.}
    \label{tab-dataset}
    \centering
    \resizebox{\linewidth}{!}{
    \begin{tabular}{l|cccc}
    \toprule
    Name & Usage & Size & Ground-truth & Rover\\
    \midrule
    \multirow{3}{*}{MIC~\cite{ding2022learning}}  & Training & 3,088 & Available & Perseverance\\
    & Validation & 386 & Available & Perseverance\\
     & Testing & 386 & Available & Perseverance\\
     \midrule
    AI4MARS~\cite{ai4mars} & Testing & 35K & N/A & \begin{tabular}[c]{@{}l@{}} Curiosity, Oppor\\-tunity, and Spirit\end{tabular}\\
    \midrule
    Mars32K~\cite{Mars32K} & Testing & 32,368 & N/A & Curiosity\\
    \bottomrule
    \end{tabular}
    }
    \vspace{-0.5em}
\end{table}

\textbf{Martian image datasets.}
Several Martian image datasets~\cite{ai4mars,Mars32K,ding2022learning} have been established through the cameras mounted on Mars rovers.
In particular, Mars32K~\cite{Mars32K} consists of 32,368 color Martian images collected by the Curiosity rover.
Moreover, a notable large-scale labeled dataset, AI4MARS~\cite{ai4mars}, collects over 35K images from the Curiosity, Opportunity, and Spirit rovers of NASA.
With crowd-sourced labels annotated by the rover planners and scientists, AI4MARS was proposed as a high-quality dataset for training and evaluating models for terrain classification on Mars.
Recently, Ding \etal~\cite{ding2022learning} have presented a Martian image dataset, \ie, MIC, which consists of high-resolution and large-scale Martian color images captured by the latest Perseverance rover~\cite{Preserverance}.
Specifically, the MIC dataset comprises 3,860 raw colorful Martian images captured by the left Mastcam-Z camera on the Perseverance rover, all at a resolution of $1152 \times 1600$.
These images are randomly divided into 3,088 images for training, 386 images for validation, and 386 images for testing.
As detailed in Table~\ref{tab-dataset},  we adopt the training, validation, and test sets of MIC as the same usage in our experiments.
Moreover, the AI4MARS and Mars32K datasets are used as our test sets, since they have no ground-truth images and can only be used for qualitative evaluation.

\textbf{Martian image compression.}
JPEG compression has been widely adopted in recent Mars exploration missions~\etal\cite{maki2020mars, PSDMSL}.
In practical applications of Martian missions, compression parameters span a broad range rather than being fixed to specific values.
As documented by Malin~\etal\cite{malin2017mars}, the Curiosity rover obtained color JPEG images with a commanded compression quality value ranging from 1 (most compression) to 100 (least compression).
Therefore, our experiments mainly focus on enhancing the quality of JPEG-compressed Martian images.
In addition, we include the High-Efficiency Video Coding with Main Still image Profile (HEVC-MSP)/Better Portable Graphics (BPG)~\cite{bpg,hevc} codec for Martian image compression, considering its prevalence in quality enhancement approaches for natural images~\cite{dcad,mfqev2,rbqe,daqe,Zheng_2022_CVPR}.

We adopt four compression settings for each codec, with the quality factor (QF) being 20, 30, 40, and 50 for JPEG, and the quantization parameter (QP) being 27, 32, 37, and 42 for BPG.
Note that these settings have also been adopted by prevalent quality enhancement approaches for natural images.

\textbf{Hyper-parameters and training.}
In the MarsQE approach, parameters $S$,  $d$, and $L$ are set to 128, 4, and 3, respectively.
All convolution operators are with a stride of 1 and a padding of 1.
To classify the input patches, we adopt an off-the-shelf Transformer-based method for Martian terrain segmentation~\cite{MarsTransformer}.
This method classifies image patches into 4 classes, and the parameter $K$ is thus set to 4.
When constructing the dictionary for reference matching, we select 6 patches for each semantic class; therefore, the parameters $\{{N_i}\}_{i=1}^{K}$ are all set to 6.
As for the loss function, we set $\lambda$ to 1.
During the training process, the Adam~\cite{adam} optimizer is applied with an initial learning rate of $10^{-4}$, beta1 of 0.9 and beta2 of 0.999.
The cosine annealing schedule~\cite{cosine-annealing} is also applied to  automatically decrease the learning rate.
We set the training batch size to 16.
A workstation with one CPU (12th Gen Intel Core i9-12900KF) and four GPUs (NVIDIA GeForce RTX 4090) is adopted in our experiments.

\subsection{Evaluation}\label{sec-sec-eval}

We evaluate the performance of our MarsQE approach on the quality enhancement of compressed Martian images.
Since there exist no quality enhancement approaches for Martian images, we compare MarsQE with widely-used approaches for natural images, including AR-CNN~\cite{ar-cnn}, DCAD~\cite{dcad}, DnCNN~\cite{dncnn}, CBDNet~\cite{cbdnet}, RBQE~\cite{rbqe}, and DAQE~\cite{daqe}.
For fair comparison, all approaches are re-trained on our training set in a non-blind manner, \ie, one model is trained for each compression setting.

\begin{table*}[!t]
  \caption{Quantitative comparison of our MarsQE and compared approaches for JPEG-compressed and BPG-compressed images over the MIC test set.
  All metrics are calculated with the raw image as the reference.
  The PSNR, PSNR-B and MUSIQ values are accurate to two decimal places, while the SSIM, MS-SSIM, and LPIPS are accurate to three decimal places.}
  \label{tab-response-efficacy}
  \centering
  \resizebox{\linewidth}{!}{
  \begin{tabular}{ l | c c c c c c | l | c c c c c c}
    \toprule
    \multicolumn{7}{c|}{JPEG} &  \multicolumn{7}{c}{BPG} \\
    \midrule
    \multirow{1}{*}{Approach} & PSNR & PSNR-B & SSIM & \scriptsize{MS-SSIM} & LPIPS & MUSIQ & \multirow{1}{*}{Approach} & PSNR & PSNR-B & SSIM & \scriptsize{MS-SSIM} & LPIPS & MUSIQ \\
    \midrule
     QF $= 20$ & 31.10 & 29.26 & 0.822 & 0.928 & 0.318 & 51.21 & 
     QP $= 42$ & 30.33 & 30.19 & 0.755 & 0.900 & 0.439 & 48.68\\ 
     AR-CNN & 32.43 & 32.14 & 0.840 & 0.948 & 0.272 & 54.72 & 
     AR-CNN & 30.67 & 30.66 & 0.758 & 0.906 & 0.383 & 49.51 \\
     DCAD & 32.75 & 32.33 & 0.847 & 0.951 & 0.252 & 55.31 & 
     DCAD & 30.77 & 30.74 & 0.759 & 0.907 & 0.374 & 49.97 \\
     DnCNN & 32.78 & 32.41 & 0.847 & 0.951 & 0.252 & 54.70 & 
     DnCNN & 30.80 & 30.76 & 0.760 & 0.907 & 0.373 & 49.59 \\       
     RBQE & 33.06 & 32.64 & 0.852 & 0.954 & 0.237 & 56.38 & 
     RBQE & 31.02 & 30.96 & 0.766 & 0.911 & 0.359 & 51.30 \\
     CBDNet & 33.14 & 32.75 & 0.854 & 0.955 & 0.230 & 57.23 & 
     CBDNet & 31.07 & 31.00 & 0.768 & 0.912 & 0.357 & 51.40 \\
     DAQE & 33.03 & 32.63 & 0.850 & 0.953 & 0.243 & 56.47 & 
     DAQE & 30.98 & 30.92 & 0.764 & 0.910 & 0.364 & 51.38 \\
     IDR & 32.88 & 32.49 & 0.850 & 0.949 & 0.248 & 55.46& 
     IDR & 30.86 & 30.81 & 0.764 & 0.910 & 0.370 & 50.34\\
     ConvIR & 33.29 & 32.91 & 0.856 & 0.956 & 0.229 & 56.99 & 
     ConvIR & 31.14 & 31.10 & 0.771 & 0.914 & 0.348 & 52.23\\
     \rowcolor{mygray}\textbf{MarsQE} & \textbf{33.36} & \textbf{32.96} & \textbf{0.857} & \textbf{0.957} & \textbf{0.223} & \textbf{57.29} & 
     \textbf{MarsQE} & \textbf{31.21} & \textbf{31.13} & \textbf{0.772} & \textbf{0.915} & \textbf{0.346} & \textbf{52.23} \\
    \midrule
    QF $= 30$ & 32.44 & 30.54 & 0.859 & 0.951 & 0.253 & 55.21 & 
    QP $= 37$ & 32.21 & 31.96 & 0.835 & 0.940 & 0.319 & 55.81\\   
    AR-CNN & 33.63 & 33.15 & 0.875 & 0.963 & 0.214 & 57.53 & 
    AR-CNN & 32.66 & 32.57 & 0.839 & 0.946 & 0.266 & 56.15 \\
    DCAD & 33.98 & 33.43 & 0.881 & 0.965 & 0.197 & 57.39 & 
    DCAD & 32.86 & 32.71 & 0.843 & 0.947 & 0.257 & 56.69 \\
    DnCNN & 34.02 & 33.45 & 0.882 & 0.965 & 0.197 & 57.23 & 
    DnCNN & 32.96 & 32.79 & 0.845 & 0.948 & 0.252 & 56.19 \\      
    RBQE & 34.28 & 33.70 & 0.886 & 0.968 & 0.182 & 58.63 & 
    RBQE & 33.21 & 33.01 & 0.850 & 0.951 & 0.237 & 57.68 \\
    CBDNet & 34.36 & 33.82 & 0.888 & 0.968 & 0.178 & 58.79 & 
    CBDNet & 33.26 & 33.06 & 0.852 & 0.951 & 0.234 & 57.67 \\ 
    DAQE & 34.28 & 33.70 & 0.885 & 0.967 & 0.188 & 58.70 & 
    DAQE & 33.20 & 33.00 & 0.849 & 0.950 & 0.244 & 57.74 \\
    IDR & 34.10 & 33.55 & 0.884 & 0.966 & 0.193 & 57.84& 
    IDR & 33.03 & 32.86 & 0.848 & 0.949 & 0.248 & 56.83\\
    ConvIR & 34.49 & 33.94 & 0.890 & 0.969 & 0.173 & 59.05 & 
    ConvIR & 33.38 & 33.18 & 0.854 & 0.952 & 0.230 & 58.05\\
    \rowcolor{mygray}\textbf{MarsQE} & \textbf{34.61} & \textbf{34.03} & \textbf{0.891} & \textbf{0.970} & \textbf{0.171} & \textbf{59.11} & 
    \textbf{MarsQE} & \textbf{33.46} & \textbf{33.25} & \textbf{0.855} & \textbf{0.953} & \textbf{0.227} & \textbf{58.19} \\
    \midrule
    QF $= 40$ & 33.24 & 31.33 & 0.878 & 0.962 & 0.211 & 56.86 & 
    QP $= 32$ & 34.36 & 33.91 & 0.896 & 0.965 & 0.202 & 59.84\\  
    AR-CNN & 34.37 & 33.80 & 0.893 & 0.971 & 0.179 & 58.88 & 
    AR-CNN & 35.00 & 34.66 & 0.903 & 0.970 & 0.157 & 60.13 \\
    DCAD & 34.82 & 34.12 & 0.901 & 0.973 & 0.160 & 58.28 & 
    DCAD & 35.42 & 34.92 & 0.909 & 0.972 & 0.151 & 60.03 \\
    DnCNN & 34.86 & 34.15 & 0.901 & 0.973 & 0.162 & 58.65 & 
    DnCNN & 35.55 & 35.02 & 0.910 & 0.972 & 0.147 & 59.67 \\    
    RBQE & 35.13 & 34.40 & 0.905 & 0.975 & 0.151 & 59.73 & 
    RBQE & 35.76 & 35.22 & 0.913 & 0.974 & 0.137 & 60.62 \\    
    CBDNet & 35.22 & 34.54 & 0.906 & 0.975 & 0.148 & 59.93 & 
    CBDNet & 35.90 & 35.44 & 0.915 & 0.975 & 0.135 & 60.85 \\
    DAQE & 35.13 & 34.41 & 0.905 & 0.974 & 0.154 & 59.81 & 
    DAQE & 35.78 & 35.28 & 0.913 & 0.974 & 0.142 & 60.69 \\
    IDR & 34.95 & 34.25 & 0.903 & 0.974 & 0.158 & 59.09& 
    IDR & 35.62 & 35.09 & 0.912 & 0.973 & 0.141 & 60.04\\
    ConvIR & 35.37 & 34.64 & 0.909 & 0.976 & 0.140 & 59.94 & 
    ConvIR & 35.99 & 35.53 & 0.916 & 0.975 & 0.133 & 60.85\\
    \rowcolor{mygray}\textbf{MarsQE} & \textbf{35.43} & \textbf{34.72} & \textbf{0.909} & \textbf{0.976} & \textbf{0.140} & \textbf{59.95} & 
    \textbf{MarsQE} & \textbf{36.07} & \textbf{35.58} & \textbf{0.917} & \textbf{0.976} & \textbf{0.131} & \textbf{60.85} \\
    \midrule
    QF $= 50$ & 33.84 & 31.94 & 0.892 & 0.968 & 0.181 & 57.99 & 
    QP $= 27$ & 36.77 & 35.95 & 0.937 & 0.981 & 0.096 & 60.69\\   
    AR-CNN & 34.96 & 34.32 & 0.906 & 0.976 & 0.154 & 59.64 & 
    AR-CNN & 37.82 & 36.93 & 0.946 & 0.985 & 0.077 & 61.30 \\
    DCAD & 35.47 & 34.67 & 0.914 & 0.977 & 0.138 & 59.21 & 
    DCAD & 38.46 & 37.29 & 0.951 & 0.986 & 0.073 & 60.82 \\
    DnCNN & 35.52 & 34.69 & 0.915 & 0.977 & 0.138 & 59.33 & 
    DnCNN & 38.55 & 37.34 & 0.952 & 0.986 & 0.071 & 60.79 \\       
    RBQE & 35.78 & 34.97 & 0.918 & 0.979 & 0.128 & 60.35 & 
    RBQE & 38.77 & 37.59 & 0.954 & 0.987 & 0.070 & 61.21 \\ 
    CBDNet & 35.92 & 35.10 & 0.920 & 0.979 & 0.122 & 60.40 & 
    CBDNet & 38.89 & 37.91 & 0.954 & 0.987 & 0.068 & 61.39 \\
    DAQE & 35.81 & 34.97 & 0.918 & 0.979 & 0.130 & 60.35 & 
    DAQE & 38.75 & 37.58 & 0.953 & 0.987 & 0.072 & 61.28 \\
    IDR & 35.60 & 34.81 & 0.916 & 0.978 & 0.135 & 59.74& 
    IDR & 38.60 & 37.44 & 0.953 & 0.986 & 0.071 & 60.90\\
    ConvIR & 36.03 & 35.17 & 0.921 & 0.980 & 0.118 & 60.43 & 
    ConvIR & 38.91 & 37.91 & 0.955 & 0.987 & 0.067 & 61.29\\
    \rowcolor{mygray}\textbf{MarsQE} & \textbf{36.09} & \textbf{35.26} & \textbf{0.921} & \textbf{0.980} & \textbf{0.119} & \textbf{60.43} & 
    \textbf{MarsQE} & \textbf{39.03} & \textbf{38.04} & \textbf{0.955} & \textbf{0.988} & \textbf{0.066} & \textbf{61.42} \\
    \bottomrule
  \end{tabular}
  }
  \vspace{-0.5em}
\end{table*}

\textbf{Quantitative performance.}
To quantify the efficacy of our and compared approaches, we measure the performance of quality enhancement in terms of several metrics, including Peak Signal-to-Noise Ratio (PSNR), Peak Signal-to-Noise Ratio including Blocking effects (PSNR-B)~\cite{psnr-b}, Structural Similarity Index Measure (SSIM)~\cite{ssim}, Multi-Scale Structural Similarity (MS-SSIM) index~\cite{ms-ssim}, Learned Perceptual Image Patch Similarity (LPIPS)~\cite{lpips}, and MUlti-Scale Image Quality transformer (MUSIQ)~\cite{musiq}.
As shown in Table~\ref{tab-response-efficacy}, our MarsQE achieves the best performance over the MIC test set in terms of all metrics.
Specifically, for the JPEG-compressed Martian images with QF $=30$, the average PSNR of MarsQE is 34.61 dB, which is 2.17 dB higher than the JPEG baseline, and 0.12 dB higher than that of the second-best approach.
In addition, the average PSNR-B is 34.03 dB, which is 3.49 dB higher than the JPEG baseline, and 0.09 dB higher than that of the second-best approach.
Similar results can be found for other QF settings and metrics.
For the BPG-compressed Martian images with QP $= 37$, the average PSNR of MarsQE is 33.46 dB, which is 1.25 dB higher than the BPG baseline, and 0.08 dB higher than that of the second-best approach.
In addition, the average PSNR-B is 33.25 dB, which is 1.29 dB higher than the BPG baseline, and 0.07 dB higher than that of the second-best approach.
Similar results can be found for other QP settings and metrics.
In summary, our MarsQE approach achieves state-of-the-art quality enhancement performance on both JPEG and BPG-compressed Martian images.

\begin{table}[!t]
  \caption{Rate-distortion performance of our MarsQE and compared approaches.
  The rate-distortion performance is measured by the BD-rate reduction (\%) with the JPEG/BPG baseline as the anchor.
  The rate is measured by the Bits Per Pixel (BPP).
  The distortion is measured by PSNR (dB) and SSIM.}
  \label{tab-rate-distortion}
  \centering
  \resizebox{\linewidth}{!}{
  \begin{tabular}{ l | cc | l | cc }
    \toprule
    \multicolumn{3}{c|}{BPP-PSNR} &  \multicolumn{3}{c}{BPP-SSIM} \\
    \midrule
    \multirow{1}{*}{Approach} & JPEG & BPG & \multirow{1}{*}{Approach} & JPEG & BPG \\
    \midrule
    AR-CNN & -24.19 & -16.14 & AR-CNN & -15.13 & -6.76 \\
    DCAD & -29.60 & -22.70 & DCAD & -19.76 & -10.82 \\
    DnCNN & -30.15 & -24.98 & DnCNN & -19.91 & -12.35 \\    
    CBDNet & -35.74 & -32.05 & CBDNet & -25.04 & -19.00 \\
    RBQE & -34.20 & -30.30 & RBQE & -23.44 & -17.45 \\
    DAQE & -33.92 & -30.19 & DAQE & -22.49 & -16.49 \\
    IDR & -32.11 & -26.70 & IDR & -21.92 & -14.22 \\
    ConvIR & -37.65 & -34.40 & ConvIR & -26.30 & -21.11 \\
    \rowcolor{mygray}\textbf{MarsQE} & \textbf{-38.94} & \textbf{-35.93} & \textbf{MarsQE} & \textbf{-27.18} & \textbf{-22.36} \\
    \bottomrule
  \end{tabular}
  }
\end{table}

\begin{figure}[!t]
  \centering
  \includegraphics[width=1.\linewidth]{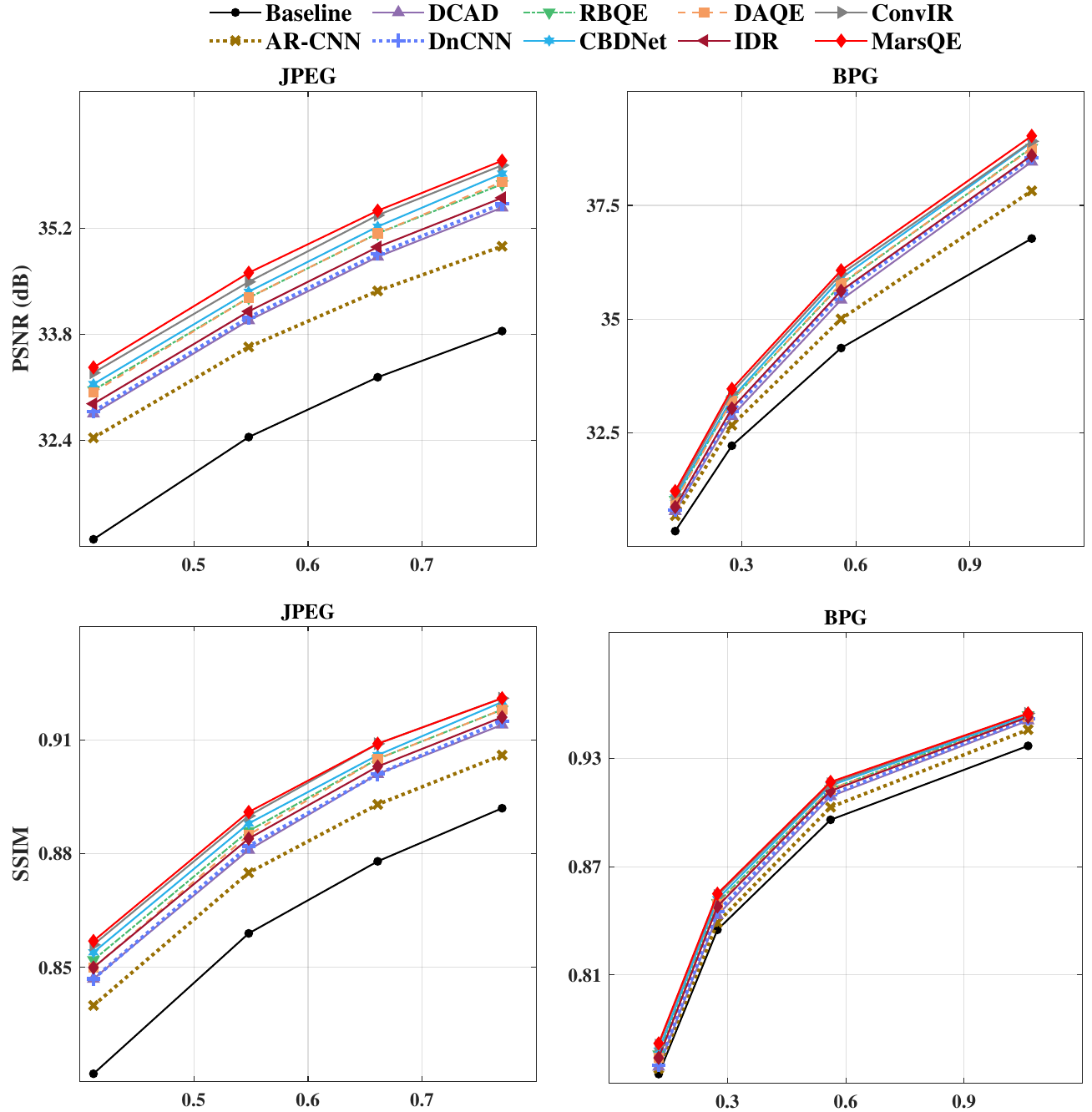}
  \caption{Rate-distortion curves of our MarsQE and compared approaches.
  The rate is measured by the bits per pixel (BPP).
  The distortion is measured by PSNR (dB) and SSIM.}
  \label{fig-rate-distortion}
  \vspace{-1em}
\end{figure}

\textbf{Rate-distortion performance.}
We further evaluate the rate-distortion performance of our MarsQE approach in Figure~\ref{fig-rate-distortion} and Table~\ref{tab-rate-distortion}.
Figure~\ref{fig-rate-distortion} shows the rate-distortion curves of different approaches over the MIC test set.
As can be seen from this figure, the rate-distortion curves of MarsQE are higher than those of other approaches, indicating a better rate-distortion performance of MarsQE.
Then, we quantify the rate-distortion performance by evaluating the reduction of Bjontegaard-rate (BD-rate)~\cite{bdrate}.
The results are presented in Table~\ref{tab-rate-distortion}.
As can be seen, for JPEG-compressed Martian images, the BD-rate reductions of our MarsQE approach on the MIC test set are averagely 38.94\% and 27.18\% with the distortion measured by PSNR and SSIM, respectively, considerably better than other approaches.
Similar results can be seen for BPG-compressed Martian images.
In summary, our MarsQE approach significantly advances the state-of-the-art rate-distortion performance for quality enhancement on Mars images.

\begin{figure*}[!t]
  \centering
  \includegraphics[width=.9\linewidth]{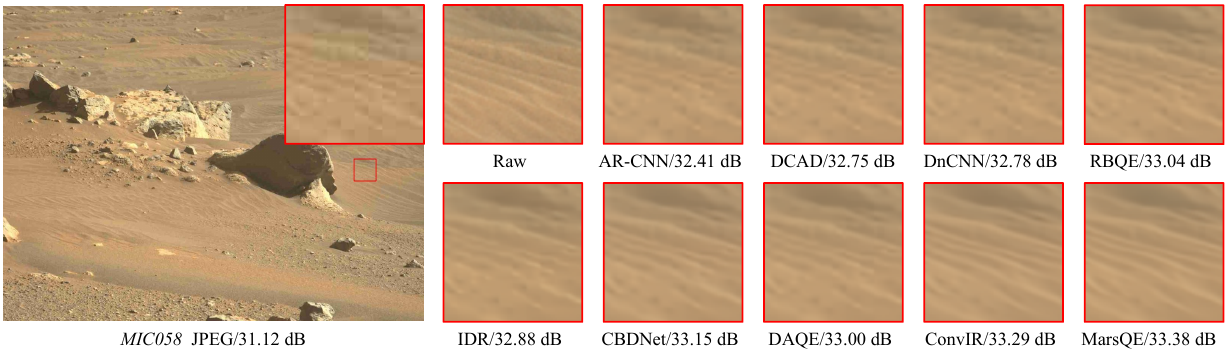}
  \caption{Qualitative comparison of our MarsQE and compared approaches.}
  \label{fig-subjective-quality}
  \vspace{-0.5em}
\end{figure*}

\begin{figure*}[!t]
  \centering
  \includegraphics[width=.9\linewidth]{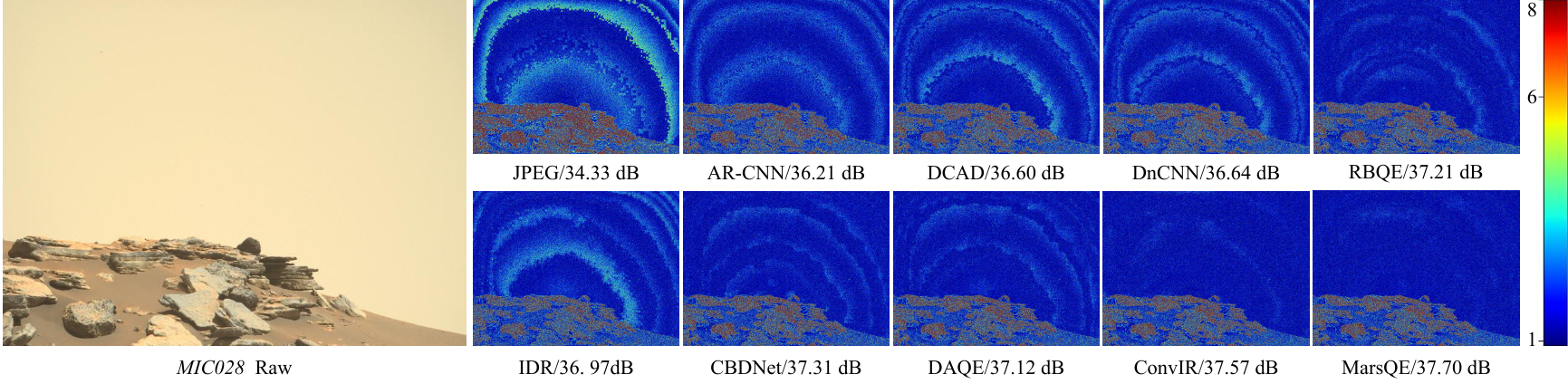}
  \caption{Residual-to-raw images of MarsQE and compared approaches. The residual values are calculated in terms of MAE.}
  \label{fig-residuals-quality}
  \vspace{-0.5em}
\end{figure*}

\textbf{Qualitative performance.}
Figure~\ref{fig-subjective-quality} compares the visual results of our MarsQE and compared approaches over the MIC test set with ground-truth images.
It can be seen in Figure~\ref{fig-subjective-quality}, our MarsQE approach successfully restores the ground-truth color and edge texture of Martian surface.
specifically, the edge of sands recovered by our MarsQE are significantly better than those recovered by other enhancement approaches.
Besides, our MarsQE approach suppresses the compression artifacts around these edges, while these artifacts can hardly be reduced by other compared approaches.
We also present the visual results of residual-to-raw images in Figure~\ref{fig-residuals-quality}. 
It is evident that our MarsQE approach effectively restores clear skies with the smallest residual values among all approaches.
To summarize, our MarsQE approaches outperforms the compared approaches in subjective quality, especially in restoring details and suppressing compression artifacts.

\begin{figure}[!t]
  \centering
  \includegraphics[width=.85\linewidth]{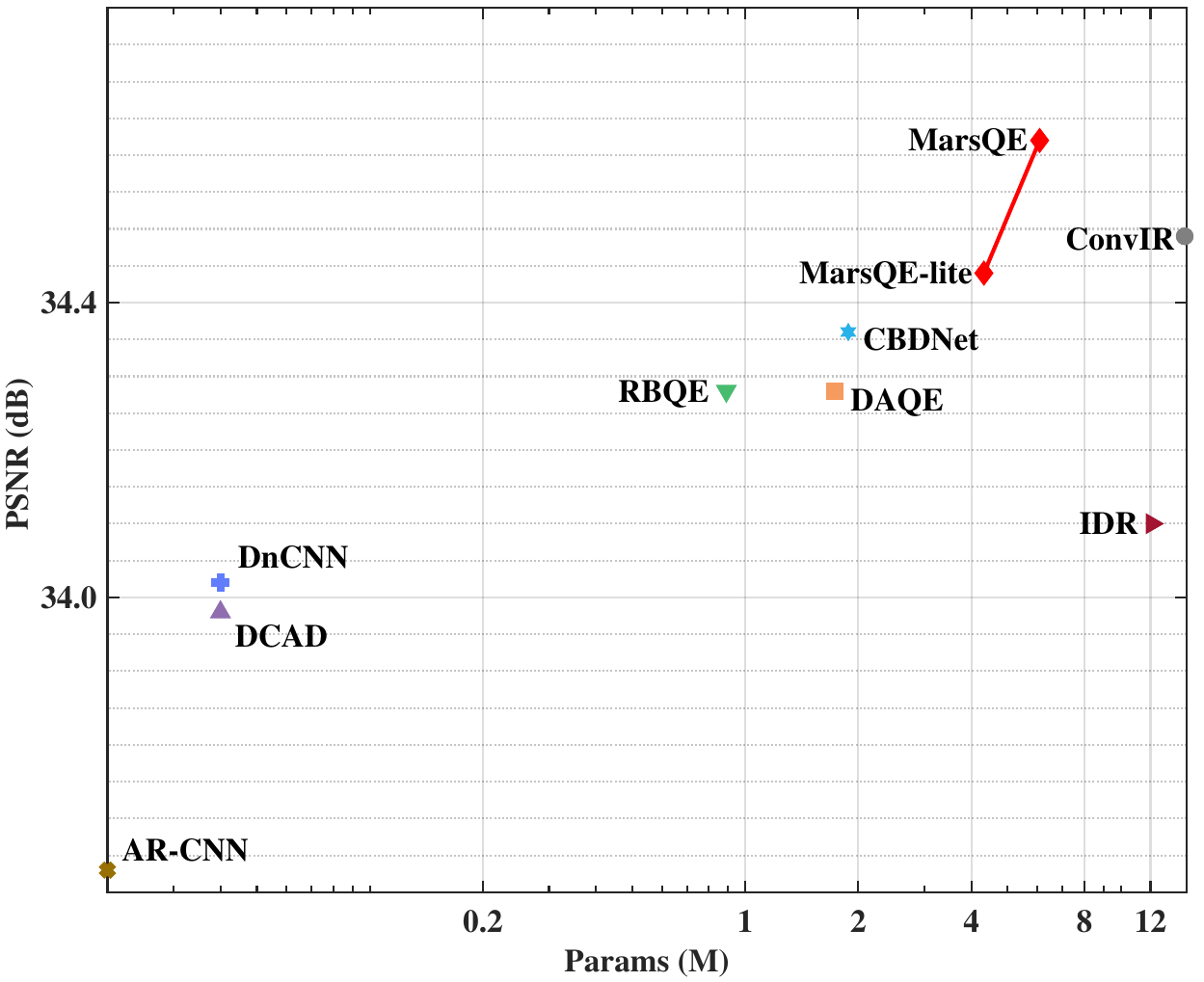}
  \caption{Parameter number and PSNR performance of the MarsQE and compared approaches over the MIC test set compressed by JPEG at QF $=30$.}
  \label{fig-efficiency}
  \vspace{-1em}
\end{figure}

\textbf{Efficiency.}
We measure the efficiency of our MarsQE approach in terms of the number of parameters and FLoating point OPerations (FLOPs).
Notably, our MarsQE approach outputs an enhanced image without the need for post-processing, as detailed in Section~\ref{sec-sec-MainNet}.
This version is referred to as MarsQE-lite.
As depicted in Figure~\ref{fig-efficiency}, utilizing only 29.1\% of the parameters of the second-best, ConvIR, MarsQE-lite is slightly behind by 0.07 dB in PSNR.
When compared with RBQE and DAQE, which have fewer parameters than ConvIR, MarsQE-lite stands out both in terms of PSNR performance and parameter efficiency.
To achieve a higher PSNR improvement, MarsQE further enhances the PSNR score by an additional 0.12 dB while having 60.0\% fewer parameters.
Although some approaches, like AR-CNN and DCAD, have fewer parameters than both our MarsQE approach and ConvIR, their PSNR performance lags by at least 0.40 dB behind MarsQE-lite and 0.59 dB behind MarsQE, respectively.
We also assess computational complexity in terms of FLOPs.
Compared to the second-best, ConvIR, our MarsQE-lite reduces FLOPs by 72.5\% with only a 0.07 dB PSNR reduction, whereas MarsQE achieves a 0.12 dB higher PSNR score with 60.0\% fewer FLOPs.
Specifically, for an input image at a resolution of $1152 \times 1600$, ConvIR requires 3.62 TMACs, while MarsQE-lite requires only 1.00 TMACs and MarsQE requires 1.45 TMACs.
In summary, our MarsQE-lite approach outperforms other compared approaches in both efficiency and enhancement performance, while our MarsQE secures higher-performance results with a modest increase in computational resources.

\subsection{Ablation Study}\label{sec-sec-ablation}

\begin{table}[!t]
  \caption{Ablation results for the modules within our MarsQE and MarsQE-lite approach in terms of PSNR (dB).}
  \label{tab-ablation-modules}
  \centering
  \resizebox{0.75\linewidth}{!}{
  \begin{tabular}{l | c}
    \toprule
     Ablation settings & PSNR(dB) \\
    \midrule
    \hspace{11pt} MarsQE &  \textbf{34.61} \\
    (A) MarsQE w/o TFM & 34.57 \\
    (B) MarsQE w/o TEM \& TFM & 34.47 \\
    \midrule
    \hspace{11pt} MarsQE-lite & \textbf{34.44} \\
    (C) MarsQE-lite w/o TFM & 34.35 \\
    (D) MarsQE-lite w/o TEM \& TFM & 34.31 \\ 
    \bottomrule
  \end{tabular}
  }
\end{table}

\begin{figure*}[!t]
  \centering
  \includegraphics[width=\linewidth]{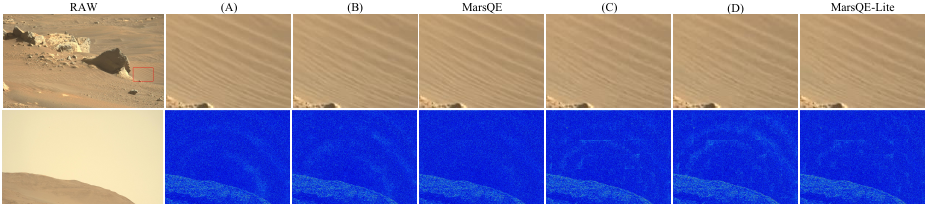}
  \caption{Qualitative comparison of our MarsQE and ablation approaches. The second row shows residual-to-raw images. The residual values are calculated in terms of MAE.}
  \label{fig-module-ablation-subjective-quality}
  \vspace{-0.8em}
\end{figure*}

\textbf{Network modules.}
In our MarsQE approach, we have introduced several crucial components.
First, TEM is designed to extract semantic-related texture patterns from reference patches.
In addition, TFM is developed to fuse semantic-related texture pattern for quality enhancement.
To assess the significance of these reference-based modules, we conduct systematic ablation studies by removing TEM and TFM from the MarsQE framework, resulting in two distinct networks labeled as (A) and (B), as illustrated in Table~\ref{tab-ablation-modules}.
Similarly, we apply a gradual module ablation strategy to the MarsQE-lite approach, generating two additional networks, denoted by (C) and (D).
These ablations involved the removal of TEM and the substitution of TFM with a simple addition operator.
Following these modifications, we re-train and evaluate all these networks using our dataset compressed by JPEG at QF $=30$.

The results, as shown in Table~\ref{tab-ablation-modules}, reveal that (1) the removal of TFM degrades PSNR by 0.04 dB for MarsQE, and (2) the further removal of TEM leads to an additional PSNR degradation of 0.10 dB.
Similar degradation can also be observed for MarsQE-lite.
We also present the visual results of ablation experiment in Figure~\ref{fig-module-ablation-subjective-quality}. 
As can be seen, our complete approach effectively suppresses the compression artifacts around these edges and restores clear skies with the smallest residual values comparing to ablated variants.
These results indicate that the reference-based modules, \ie, TEM and TFM, positively contribute to the overall performance of image quality enhancement, underscoring their importance in our approach.

\begin{figure}[!t]
  \centering
  \includegraphics[width=\linewidth]{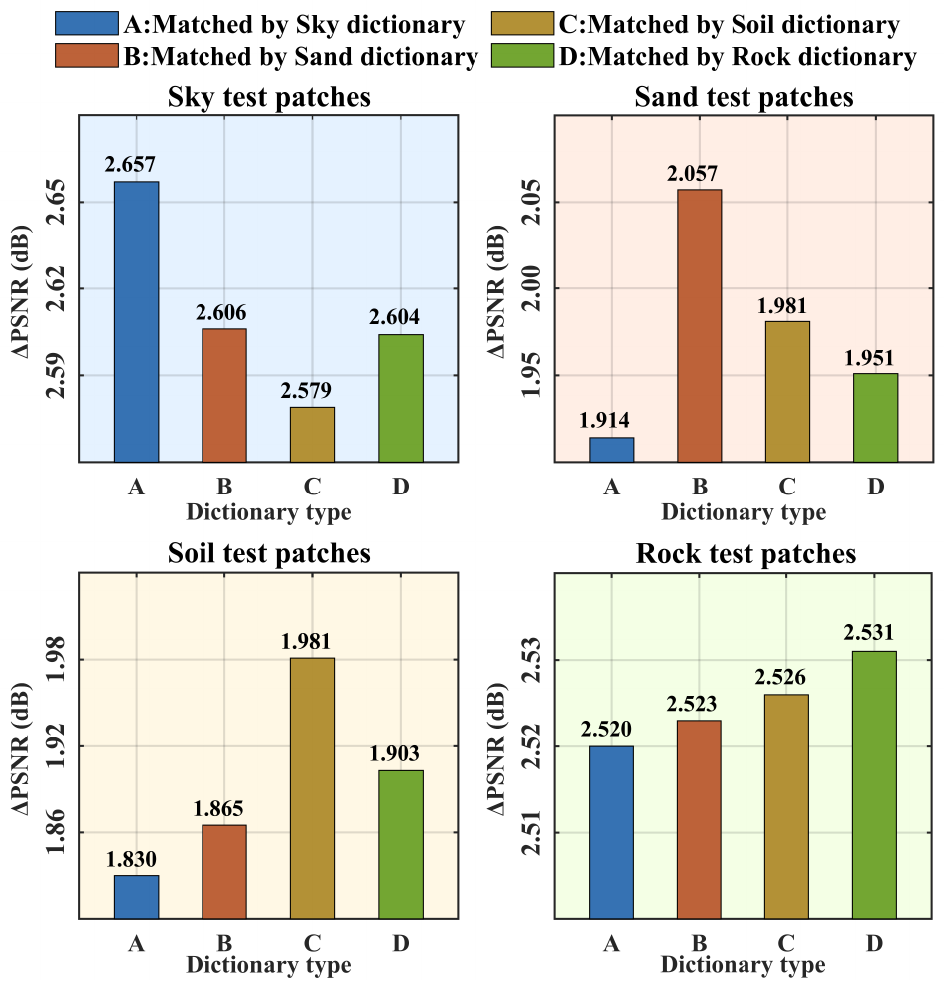}
  \caption{Statistics of the improved PSNR ($\Delta$PSNR) for four semantic classes of test patches.}
  \label{fig-ablation-semantics}
  \vspace{-1em}
\end{figure}

\begin{figure}[!t]
  \centering
  \includegraphics[width=.83\linewidth]{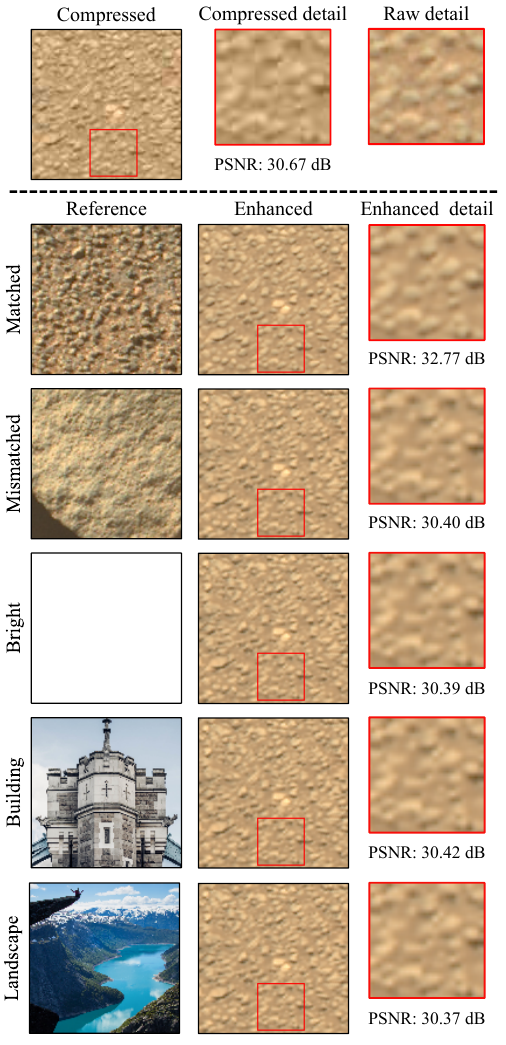}
  \caption{Qualitative results of our MarsQE approach when using texture-distinct reference patches.}
  \label{fig-ablation-mismatched-patches}
  \vspace{-1em}
\end{figure}

\begin{table}[!t]
  \caption{Ablation results for the patch number of reference dictionary within our MarsQE approach in terms of PSNR, PSNR-B, and LPIPS.}
  \label{tab-patchnum-ablation}
  \centering
  \resizebox{.8\linewidth}{!}{
  \begin{tabular}{l | ccccc}
    \toprule
    Patch number & 1 & 3 & 6 & 9 & 12 \\
    \midrule
    PSNR & 34.52 & 34.56 & 34.61 & 34.62 & 34.62 \\
    PSNR-B & 33.92 & 33.98 & 34.03 & 34.04 & 34.04 \\
    LPIPS & 0.175 & 0.173 & 0.171 & 0.171 & 0.171 \\
    \bottomrule
  \end{tabular}
  }
  \vspace{-0.5em}
\end{table}

\textbf{Patch number in reference dictionary}.
During the construction of the reference dictionaries, SMM selects $N_k$ patches for k-th semantic class to construct a reference dictionary $\{\mathbf{R}^{k}_{i}\}_{i=1}^{N_k}$.
To verify the effect of $\{\mathbf{R}^{k}_{i}\}_{i=1}^{N_k}$ value on quality enhancement performance, we conduct multiple experiments with different $\{{N_i}\}_{i=1}^{K}$ values, where $\{{N_i}\}_{i=1}^{K}$ are all set to {3, 6, 9, 12}.
Specifically, during the reference dictionary construction, we adopt a widely used clustering metric, \ie, Calinski-Harabasz (CH) index~\cite{CHIdex}, to find the best $\{{N_i}\}_{i=1}^{K}$.
Note that higher CH index indicates better cluster performance.
The corresponding CH index values were 4835.81, 5628.02, 5471.60, and 5387.31, respectively, demonstrating optimal coverage at $\{{N_i}\}_{i=1}^{K}=6$.

Furthermore, we retrain and test our MarsQE approach of different $\{{N_i}\}_{i=1}^{K}$ values.
As shown in Table~\ref{tab-patchnum-ablation}, Within the certain range ($\{{N_i}\}_{i=1}^{K} \leq 6$), the CH index gets better as $\{{N_i}\}_{i=1}^{K}$ increases.
When $\{{N_i}\}_{i=1}^{K}$ increases from 6 to 12, the improvement of PSNR is small, indicating that the number of patches in dictionary reaches saturation.
Since the model($\{{N_i}\}_{i=1}^{K}$ = 6) performs almost on par with model($\{{N_i}\}_{i=1}^{K}$ = 12) but with a smaller dictionary, $\{{N_i}\}_{i=1}^{K}$ are all set to 6 in our MarsQE.

\textbf{Semantics-based reference.}
During the construction of the reference dictionaries, we introduce four semantically distinct reference dictionaries for MarsQE, encompassing sky, sand, soil, and rock reference patches.
To validate the effectiveness of these dictionaries, we adopt each dictionary independently for reference, rather than using all dictionaries collectively.
For fair comparison, all validations are performed on our test set compressed by JPEG at QF $=30$.
The exclusive use of sky, sand, soil, and rock reference dictionaries results in a degradation of PSNR values over our test set from 34.61 to 34.53, 34.57, 34.59, and 34.56 dB, respectively.
Moreover, we evaluate the PSNR results for each semantic class of patches.
As presented in Figure~\ref{fig-ablation-semantics}, when the semantic class of patches is consistent with that of dictionary, the PSNR values achieve their highest.
The PSNR degradation of semantically mismatched references reaches up to 0.15 dB when sky reference patches are provided to soil test patches (\ie, from 1.98 to 1.83 dB).
We also provide visual results with different reference in Figure~\ref{fig-ablation-mismatched-patches}.
It can be seen that using reference patches with differing semantics can significantly impair the restoration of color and edges in rocks.
In summary, the semantics-based reference pose a positive effect on our semantic-informed quality enhancement in terms of both quantitative and qualitative performance.

\begin{figure}[!t]
  \centering
  \includegraphics[width=.95\linewidth]{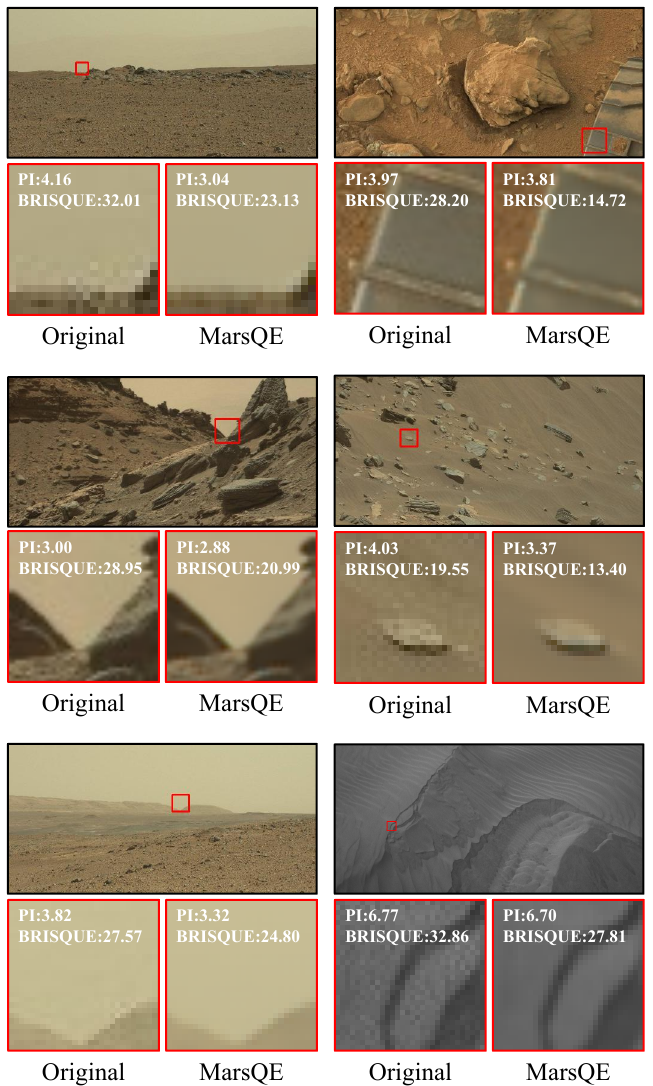}
  \caption{Qualitative results of our MarsQE approach on the Mars32K and AI4MARS datasets without fine-tuning.
  }
  \label{fig-generalization}
  \vspace{-1.5em}
\end{figure}

\subsection{Generalization Capability Evaluation}\label{sec-sec-Generalization}

We further validate our MarsQE approach on the datasets beyond the MIC dataset~\cite{ding2022learning}, \ie, AI4MARS~\cite{ai4mars} and Mars32K~\cite{Mars32K}, without fine-tuning the MarsQE network.
Recall that MarsQE is trained on the MIC dataset with ground-truth RAW images captured by the Perseverance rover.
In comparison, the AI4MARS and Mars32K datasets contain only compressed images captured by Curiosity, Opportunity, and Spirit rovers, without ground-truth counterparts.
Although JPEG is used in both datasets, aligning with our Section~\uppercase\expandafter{\romannumeral5}-B experiments, this expanded evaluation presents a significant challenge to MarsQE due to unseen data lacking ground-truth references and featuring unknown compression parameters.
Notably, in AI4MARS, 25.7\% of the images, captured by Curiosity's Navigation camera (Navcam), are grey-scale, posing a further challenge due to the absence of color information.

To be specific, we re-train a single model and evaluate it on mixed JPEG bitstreams with QF ranging from 20 to 50.
The PSNR result on the test set under mixed QF $=20–50$ bitstreams is 34.65 dB, demonstrating effective blind enhancement across all settings.
Additionally, we test the pre-trained MarsQE model on images from Mars32K and AI4MARS.
Given the absence of ground-truth data, we present the qualitative results and no-reference quality assessment results of MarsQE.
The qualitative results, as illustrated in Figure~\ref{fig-generalization}, underscore MarsQE's proficiency in suppressing compression artifacts, particularly the ringing effect observed at the boundaries of stones, sand, and rovers.
In addition, we evaluate the quality of these images in terms of two widely-used no-reference quality metrics, \ie, Perceptual Index (PI)~\cite{PI} and Blind/Referenceless Image Spatial Quality Evaluator (BRISQUE)~\cite{brisque}.
Note that smaller PI and BRISQUE signify a higher degree of perceptual quality.
As shown in Figure~\ref{fig-generalization}, both metrics demonstrates the effectiveness of MarsQE on enhancing unseen compressed images with improved quality.
To summarize, the MarsQE approach not only showcases its adaptability and robustness through these assessments but also confirms the potential applicability across a spectrum of Martian exploration missions, evidencing a broad generalization capability of our MarsQE approach.

\subsection{Discussion and Limitations}\label{sec-sec-Discussion}

Compared with previous quality enhancement works, the strength of our MarsQE approach lies in its outstanding performance within the specific target scenario.
However, everything has two sides, and MarsQE approach also has its weakness.
We summarize the weaknesses in two points as follows:
(1) The limitation for scene: 
Despite our approach having strong generalization capabilities, it is still difficult to handle some complex cases.
For example, as Martian exploration continues, new terrain types may emerge that differ significantly from those in the existing reference dictionary. 
In such cases, the reference-based modules fail to extract meaningful texture features from the reference database.
(2) The limitation for task: 
Currently, MarsQE treats terrain segmentation and quality enhancement as separate tasks.
However, these tasks could be mutually beneficial: enhanced details could improve segmentation accuracy, while semantic information from segmentation could guide more effective pixel-level restoration.
Regarding future research directions, MarsQE could be extended by integrating a dynamic reference dictionary and jointly training both tasks to enhance generalization and overall performance.

\section{Conclusion}
In this paper, we have proposed the MarsQE approach for enhancing the quality of compressed Martian images.
Specifically, we first obtained two main findings about Martian image by analyzing over a large-scale Martian image dataset, \ie, a notable similarity both within and between images, and compact texture representation characterized by a limited number of semantic classes.
Inspired by these findings, we developed the framework of our MarsQE approach with two phases: semantic-based reference matching and semantic-informed quality enhancement.
The semantic-based reference matching phase is powered by SMM, which identifies reference images with similar textures based on their semantic content.
In the phase of semantic-informed quality enhancement, we implemented TEM and TFM to extract and integrate transferable features, respectively.
Additionally, we introduced a post-enhancement network that aims at reducing the blocking artifacts caused by patch-wise enhancement.
The experimental results demonstrated the superior performance of our MarsQE approach on the quality enhancement of Martian images.

\ifCLASSOPTIONcaptionsoff
  \newpage
\fi

\bibliographystyle{IEEEtran}
\bibliography{refs}

\vspace{-2em}
\begin{IEEEbiography}[{\includegraphics[width=1in,height=1.25in,clip,keepaspectratio]{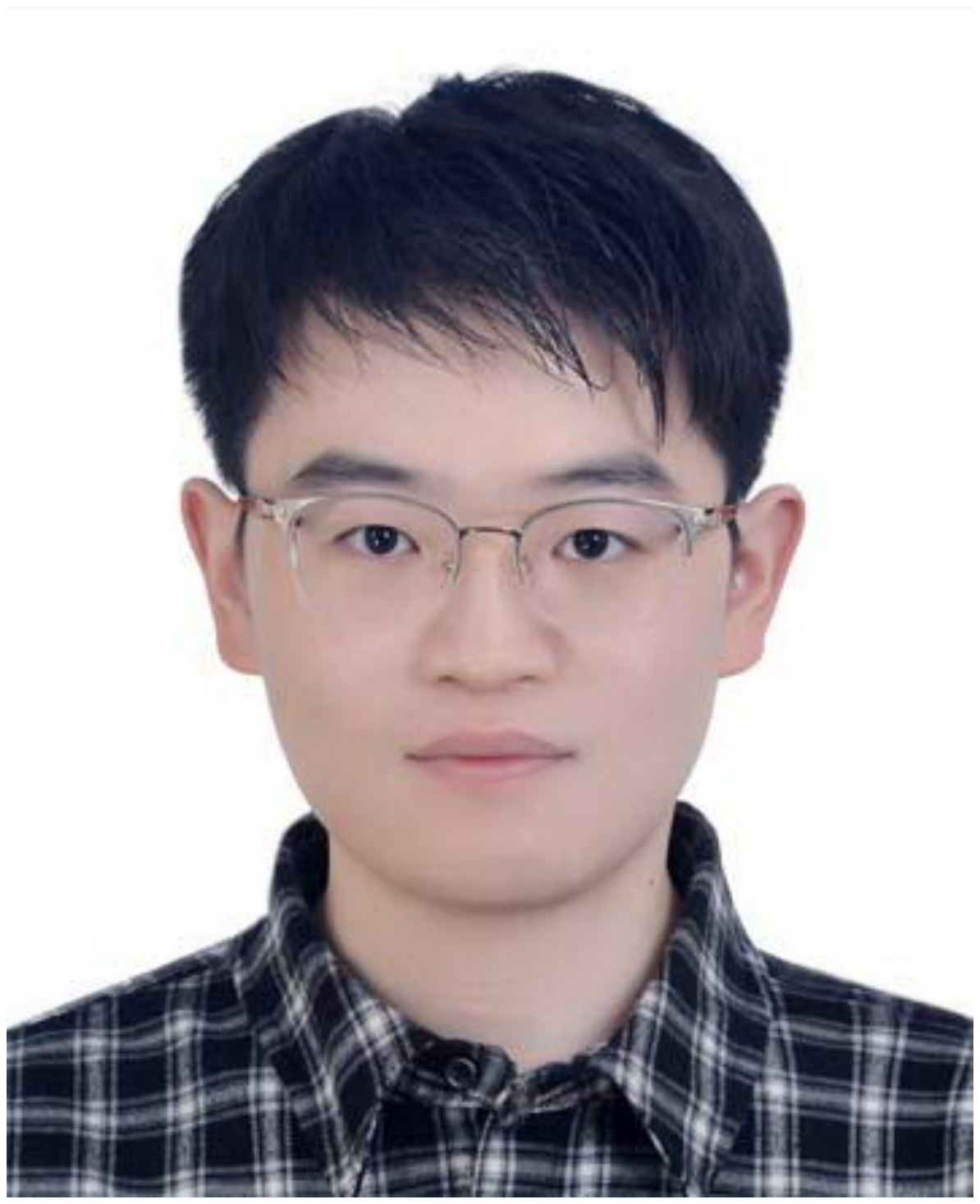}}]{Chengfeng Liu}
received the M.S degree from Beihang University in 2025. He previously received his B.S degree from Beihang University in 2022. His research interests mainly include image and video processing, quality enhancement, and computer vision.
\end{IEEEbiography}

\begin{IEEEbiography}
[{\includegraphics[width=1in,height=1.25in,clip,keepaspectratio] {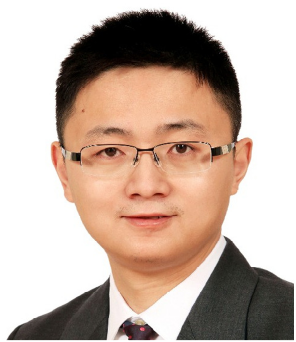}}]{Mai Xu} (Senior Member, IEEE)
received the B.S. degree from Beihang University, Beijing, China, in 2003, the M.S. degree from Tsinghua University, Beijing, China, in 2006, and the Ph.D. degree from Imperial College London, London, U.K., in 2010. From 2010 to 2012, he was a Research Fellow with the Department of Electrical Engineering, Tsinghua University. Since January 2013, he has been with Beihang University, where he was an Associate Professor and promoted to Full Professor in 2019. During 2014 to 2015, he was a Visiting Researcher of MSRA. His research interests mainly include image processing and computer vision. He has authored or coauthored more than 100 technical papers in international journals and conference proceedings, e.g., IJCV, IEEE TPAMI, TIP, J-STSP, CVPR, ICCV, ECCV, and AAAI. He is the recipient of best paper awards of three IEEE conferences. He served as an Associate Editor of IEEE TIP, a Lead Guest Editor of IEEE J-STSP, and an Area Chair or TPC Member of many conferences, such as ICME, AAAI, \etc. He is an elected member of Multimedia Signal Processing Technical Committee, IEEE Signal Processing Society.
\end{IEEEbiography}

\begin{IEEEbiography}[{\includegraphics[width=1in,height=1.25in,clip,keepaspectratio]{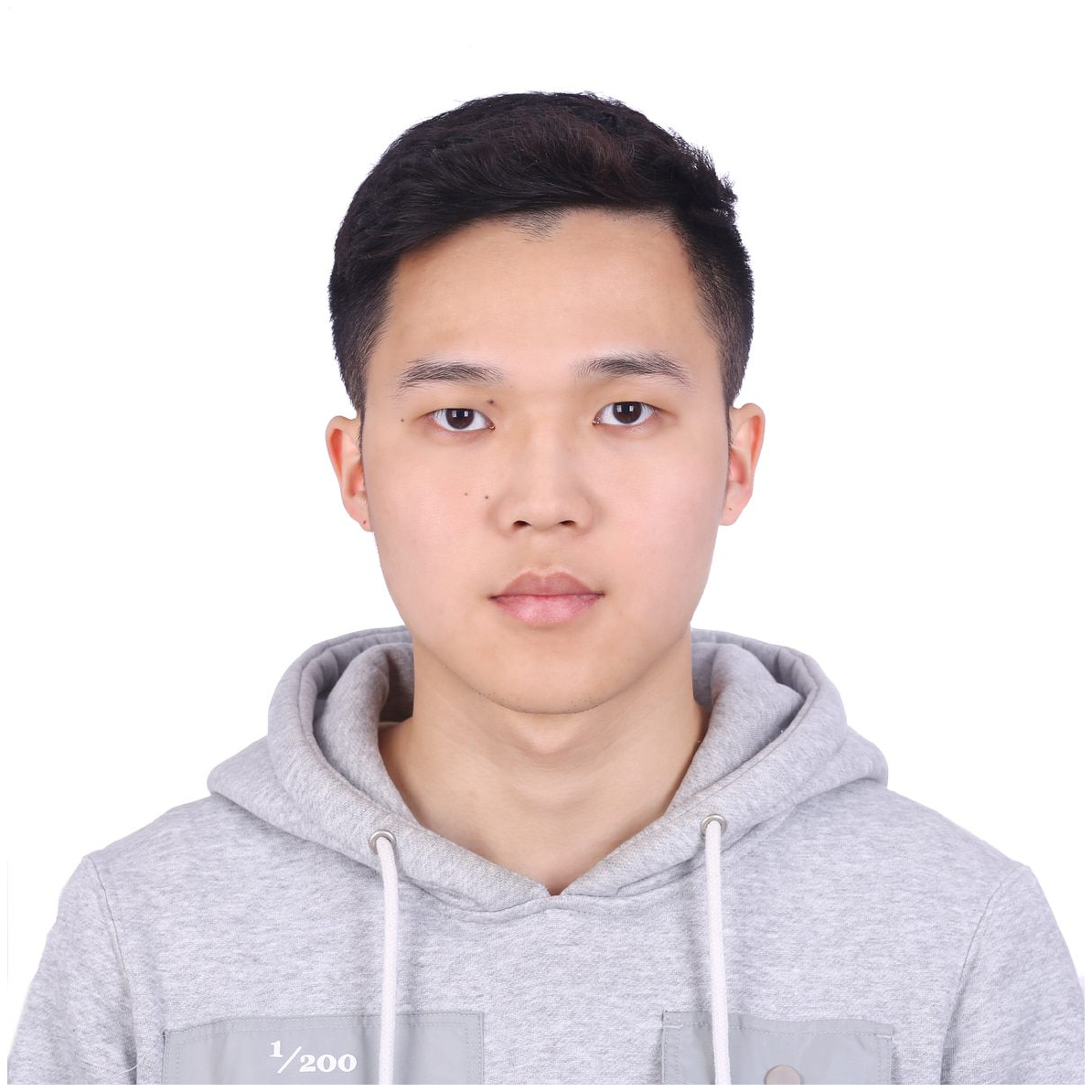}}]{Qunliang Xing}
received his Ph.D. in Engineering from Beihang University in 2025, where he was advised by Professor Mai Xu. He also completed a university-level honors program during his Ph.D. studies. He previously received his B.S degree from Beihang University in 2019 and completed the undergraduate honors program.
His research interests include Computer Vision and Multimedia. He has published several papers in top journals and conferences such as IEEE TPAMI and CVPR, including highly cited papers recognized by the Essential Science Indicators (ESI). His publications have received over 600 citations. He has won two NTIRE championships. His open-source projects have garnered more than 800 stars. He was selected for the Tencent Rhino-bird Open-source Training Program and recognized as an OpenMMLab Active Contributor. He serves as a reviewer for CVPR, TIP, and other venues.
\end{IEEEbiography}

\begin{IEEEbiography}[{\includegraphics[width=1in,height=1.25in,clip,keepaspectratio]{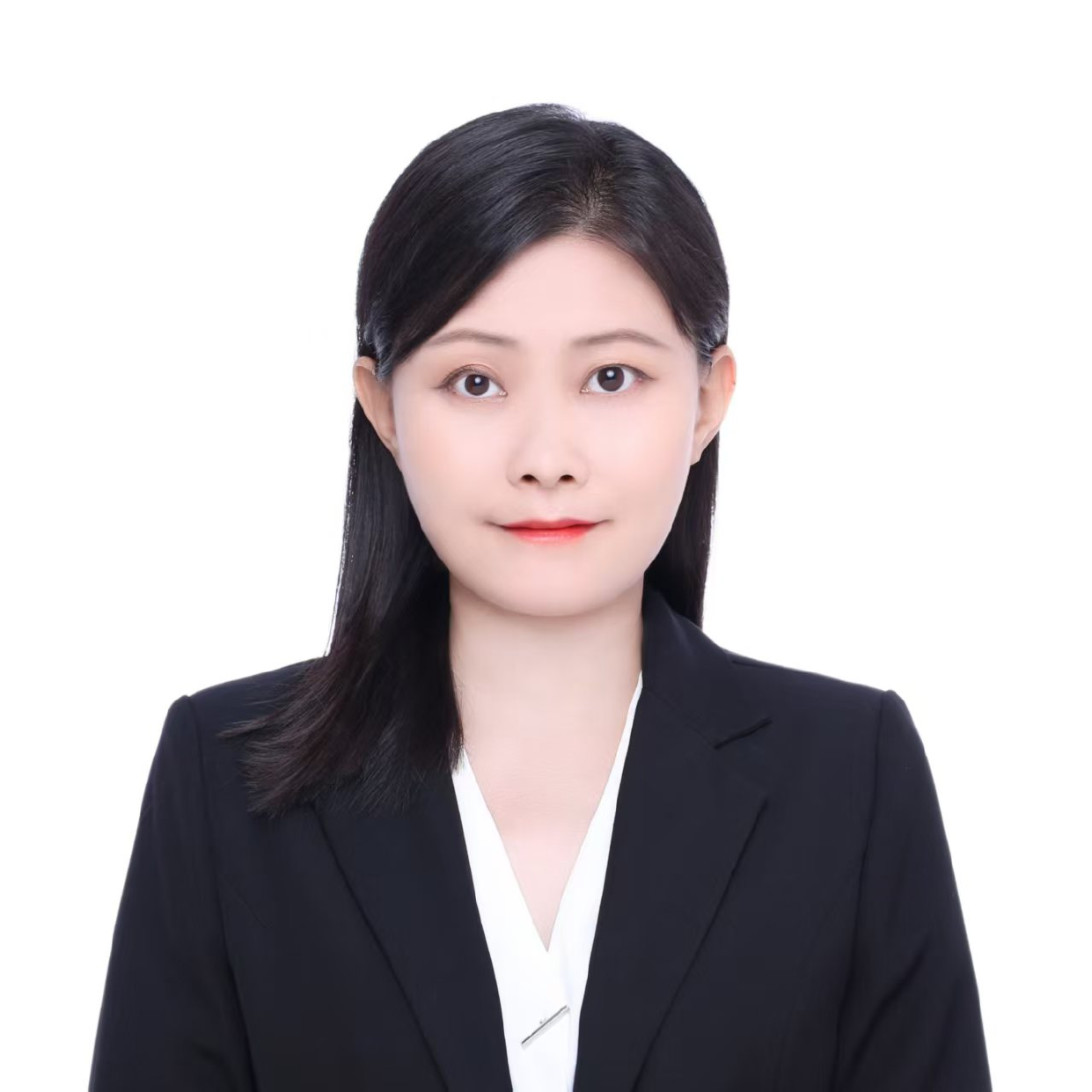}}]{Xin Zou}
received the M.S degree from Changchun University of Science and Technology in Jun. 2009. She is currently pursuing the Ph.D. degree with the School of Electronic and Information Engineering, Beihang University, Beijing, China. Her research interests mainly include spacecraft design, image processing and computer vision.
\end{IEEEbiography}

\vfill

\end{document}